\documentclass[fleqn,usenatbib]{mnras}
\pdfoutput=1

\usepackage{txfonts}

\usepackage[T1]{fontenc}
\usepackage{ae,aecompl}

\usepackage{graphicx}	
\usepackage{amsmath}	
\usepackage{amssymb}	
\usepackage{comment}
\usepackage{relsize}	
\usepackage{xspace}
\usepackage{subfig}     

\newcommand{\Ms}{\ensuremath{\textrm{M}_{\odot}}\xspace}

\newcommand{\Mj}{\ensuremath{\textrm{M}_{\textrm{J}}}\xspace}
\newcommand{\Rs}{\ensuremath{\textrm{R}_{\odot}}\xspace}

\newcommand{\e}[1]{\ensuremath{\times 10^{#1}}}
\newcommand{\numax}{\ensuremath{\nu_{\textrm{max}}}\xspace}
\newcommand{\delnu}{\ensuremath{\Delta\nu}\xspace}
\newcommand{\dpi}{\ensuremath{\Delta\Pi}\xspace}
\newcommand{\teff}{\ensuremath{T_{\textrm{eff}}}\xspace}
\newcommand{\logg}{\ensuremath{\log{g}}\xspace}
\newcommand{\feh}{\ensuremath{\textrm{[Fe/H]}}\xspace}
\newcommand{\mh}{\ensuremath{\textrm{[m/H]}}\xspace}

\newcommand{\Kepler}{\ensuremath{\emph{Kepler}}\xspace}
\newcommand{\tabgap}{\noalign{\vskip 1mm}}

\title[Red-giant+M-dwarf eclipsing binary]{KOI-3890: A high mass-ratio asteroseismic red-giant$+$M-dwarf eclipsing binary undergoing heartbeat tidal interactions}

\author[J. S. Kuszlewicz et al.]{
James S. Kuszlewicz,$^{1,2,3}$\thanks{E-mail: kuszlewicz@mps.mpg.de}
Thomas S. H. North,$^{2,3}$
William J. Chaplin,$^{2,3}$
\newauthor
Allyson Bieryla,$^{4}$
David W. Latham,$^{4}$
Andrea Miglio,$^{2,3}$
Keaton J. Bell,$^{1,3}$
\newauthor
Guy R. Davies,$^{2,3}$
Saskia Hekker,$^{1,3}$
Tiago L. Campante,$^{5,6}$
Sebastien Deheuvels,$^{7,8}$
\newauthor
and Mikkel N. Lund$^{3,2}$
\\
$^{1}$Max-Planck-Institut f\"{u}r Sonnensystemforschung, Justus-von-Liebig-Weg 3, 37077 G\"{o}ttingen, Germany \\
$^{2}$School of Physics and Astronomy, University of Birmingham, Edgbaston, Birmingham, B15 2TT, UK\\
$^{3}$Stellar Astrophysics Centre (SAC), Department of Physics and Astronomy, Aarhus University, Ny Mukegade 120, \\ DK-8000 Aarhus C, Denmark \\
$^{4}$Harvard-Smithsonian Center for Astrophysics, Cambridge, Massachusetts 02138, USA \\
$^{5}$Instituto de Astrof\'{i}sica e Ci\^{e}ncias do Espa\c{c}o, Universidade do Porto, Rua das Estrelas, P-4150-762 Porto, Portugal\\
$^{6}$Departamento de F\'{i}sica e Astronomia, Faculdade de Ci\^{e}ncias da Universidade do Porto, Rua do Campo Alegre, s/n, P-4169-007 Porto, Portugal\\
$^{7}$Universit\'{e} de Toulouse, UPS-OMP, IRAP, Toulouse, France \\
$^{8}$CNRS, IRAP, 14, avenue Edouard Belin, F-31400 Toulouse, France \\
}

\date{Accepted XXX. Received YYY; in original form ZZZ}

\pubyear{2018}

\begin{document}
\label{firstpage}
\pagerange{\pageref{firstpage}--\pageref{lastpage}}
\maketitle
\begin{abstract}
KOI-3890 is a highly eccentric, 153-day period eclipsing, single-lined spectroscopic binary system containing a red-giant star showing solar-like oscillations alongside tidal interactions. The combination of transit photometry, radial velocity observations, and asteroseismology have enabled the detailed characterisation of both the red-giant primary and the M-dwarf companion, along with the tidal interaction and the geometry of the system. The stellar parameters of the red-giant primary are determined through the use of asteroseismology and grid-based modelling to give a mass and radius of $M_{\star}=1.04\pm0.06 \Ms$ and $R_{\star}=5.8\pm0.2 \Rs$ respectively. When combined with transit photometry the M-dwarf companion is found to have a mass and radius of $M_{\mathrm{c}}=0.23\pm0.01 \Ms$ and $R_{\mathrm{c}}=0.256\pm0.007 \Rs$. Moreover, through asteroseismology we constrain the age of the system through the red-giant primary to be $9.1^{+2.4}_{-1.7}\mathrm{Gyr}$. This provides a constraint on the age of the M-dwarf secondary, which is difficult to do for other M-dwarf binary systems. In addition, the asteroseismic analysis yields an estimate of the inclination angle of the rotation axis of the red-giant star of $i=87.6^{+2.4}_{-1.2}$ degrees. The obliquity of the system\textemdash the angle between the stellar rotation axis and the angle normal to the orbital plane\textemdash is also derived to give $\psi=4.2^{+2.1}_{-4.2}$ degrees showing that the system is consistent with alignment. We observe no radius inflation in the M-dwarf companion when compared to current low-mass stellar models.

\end{abstract}





\section{Introduction}
Over the past decade, there has been a space-based revolution courtesy of the \emph{Kepler} \citep{2010Borucki_kepler} and CoRoT \citep{2006Baglin_corot} missions. The study of solar-like oscillations\textemdash those excited and damped by near-surface turbulent convection\textemdash had long been confined to the Sun save a few rare cases; for example: Arcturus \citep{1987ApJ...317L..79S}, Procyon \citep{1991ApJ...368..599B, 2008ApJ...687.1180A, 2010ApJ...713..935B}, and $\alpha$CenB \citep{2005ApJ...635.1281K}. The long-baseline, high-quality datasets available as a result of these space missions have resulted in a golden-age for the field, with oscillations detected in hundreds of solar-type stars (e.g. \citealt{2014ApJS..210....1C}) and thousands of red giants (e.g. \citealt{2009A&A...506..465H,2009Natur.459..398D,2011MNRAS.414.2594H,2016ApJ...827...50M,2018ApJS..236...42Y}).

Eclipsing binary systems in which the primary component is a red-giant star showing solar-like oscillations are relatively rare. Asteroseismology gives an unprecedented opportunity to better characterise these systems \citep{2010ApJ...713L.187H,2013A&A...556A.138F,2013Gaulme,2014ApJ...785....5G,2016Gaulme,2014A&A...564A..36B,2014Miglio,2016AN....337..793B,2016ApJ...818..108R,2017sf2a.conf...89B,2018MNRAS.476.3729B,2018MNRAS.tmp.1060T}. The detection and characterisation of stellar oscillations not only allows inferences to be made about the internal processes of stars \citep{2012Mosser_spindown,2013Mont,2017Eggenberger,2018A&A...610A..80H}, they also allow access to other properties of the star such as ages (e.g. \citealt{2010ARA&A..48..581S,2013MNRAS.429..423M,2018MNRAS.475.5487S}) and the stellar inclination angle with respect to our line of sight (e.g. \citealt {Kepler56,2014A&A...570A..54L,2016ApJ...819...85C,2018arXiv180507044K}; Kuszlewicz et al. submitted and references therein).

Amongst the many stars observed with \emph{Kepler}, a new class of eccentric, ellipsoidal binary systems was discovered that showed photometric tidal distortions \citep{2011Welsh, 2012ApJ...753...86T}. The distinct, ellipsoidal variations seen in the light curve are a result of the large tidal distortions of the surface of the star during periastron which leads to their name, ``heartbeat" stars.  The theory behind these tidal distortions was derived by \cite{1995Kumar} and it was shown that the morphology of these distortions can lead to constraints on the inclination of the orbit, eccentricity, and argument of periastron without the need for an eclipse to be observed.

It is possible to find heartbeat systems where one component is an oscillating red giant. \cite{2014A&A...564A..36B} analysed 18 such systems where the primary star was a red giant showing solar-like oscillations, of which only three were eclipsing and two had determined mass fractions. Heartbeat systems that show solar-like oscillations are of considerable value because, when combined with possible radial velocity data, they can be characterised in great detail. Since they provide constraints on the eccentricity and inclination of the orbit, even when the system is not eclipsing these important parameters can be derived. If the systems are also eclipsing this can help improve constraints even more. This is particularly the case as there is a tendency for heartbeat stars to have low-mass companions, and so obtaining radial velocities for each component in the system is difficult. Having another means to infer the properties of the primary star, such as asteroseismology, is invaluable since this can propagate through to tighter constraints on the secondary star.

In addition to observing hearbeat systems where one component was an oscillating red giant, one star (KIC 5006817) in the sample of \cite{2014A&A...564A..36B} was identified as being a red-giant+M-dwarf binary system. However unlike the system in this work, it was not an eclipsing system and the orbital parameters of the system were instead inferred from the heartbeat signal. \cite{2016Gaulme} had three such red-giant+M-dwarf systems in their sample, all of which were eclipsing, and used the asteroseismic scaling relations to derived the physical properties of the primary and secondary components. In this work, we are the first to apply grid-based modelling to these types of red-giant+M-dwarf systems with an oscillating component that are both eclipsing and show heartbeat signals.

Precise constraints on the inclination angle of the orbit in an eclipsing (or transiting) system through the analysis of the heartbeat signal can be combined with knowledge about the inclination angle of the star (through, for example, asteroseismology; \citealt{1985ApJ...292..238P, Gizon2003}) to constrain the obliquity of the system\footnote{The true obliquity (as opposed to the sky-projected obliquity), $\psi$, is the angle between the binary orbital axis and the stellar rotation axis.}. The obliquity is an important parameter for understanding binary systems because it can help shed light on formation mechanisms (e.g. \citealt{1992ApJ...400..579B, 2010MNRAS.401.1505B}) and the dynamical evolution of the systems \citep{1979A&A....77..145M}. The most extensive investigation into the obliquities of eclipsing binary stars has been performed by the BANANAs project \citep{2007A&A...474..565A, 2009Natur.461..373A, 2011ApJ...726...68A, 2013ApJ...767...32A, 2014ApJ...785...83A}. During these investigations, five systems were studied using the Rossiter-MacLaughlin effect (see \citealt{2007ASPC..366..170W} or \citealt{2007ApJ...655..550G} for an overview) to derive the sky-projected obliquity\footnote{The sky-projected obliquity, $\lambda$, is defined as the angle between the projections of the orbital and rotation axes on the sky.} of each star in the binary. All of the binaries analysed were close-in, with orbital periods in the range of $\sim6-16$ days, of which two are misaligned (CV Vetorum and DI Herculis) and the others are consistent with alignment. Heartbeat systems, on the other hand, can have orbital periods larger than $\sim$100 days, and their companions tend to be low-mass (and so they are likely to be single-lined spectroscopic systems), any measurement of the Rossiter-MacLaughlin effect would be very difficult, if not unobtainable with current instruments. Therefore, to determine the obliquity, the system needs to be eclipsing and the inclination angle of the star must have a measurement. In the case where the primary is an oscillating red-giant this becomes possible through asteroseismology.

In this work we aim to derive the properties of the components of an eclipsing binary system through the use of asteroseismology, eclipse fitting and radial velocity analysis. In addition, we aim to constrain the geometry of the system by inferring the inclination angle of the red giant primary to then obtain the obliquity of the system, which gives information as to whether the system is aligned.

\begin{figure*}
     \centering
     \subfloat[\label{fig:3890phase}]{\includegraphics[width=\columnwidth]{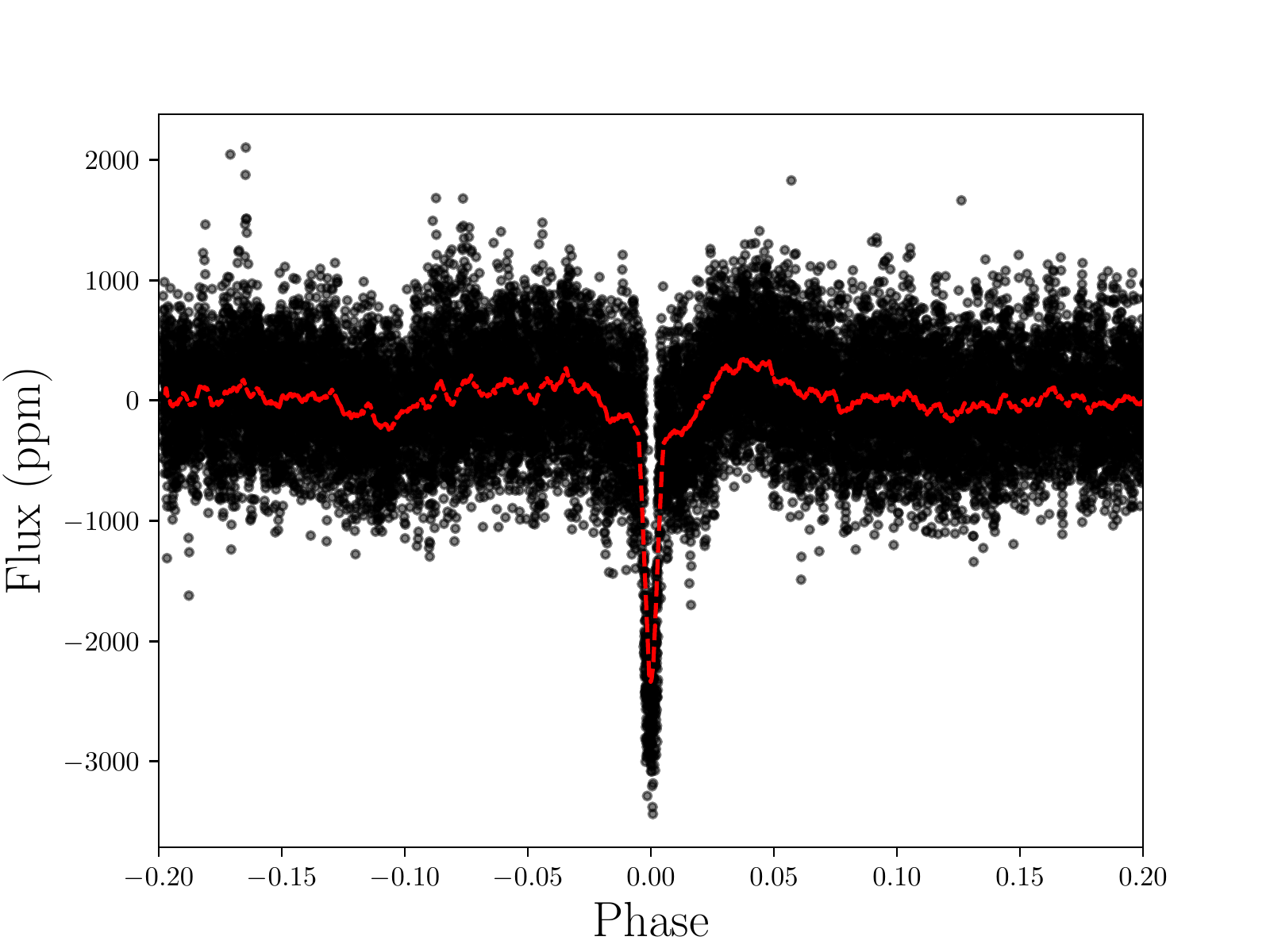}}
     \hfill
     \subfloat[\label{fig:3890tran}]{\includegraphics[width=\columnwidth]{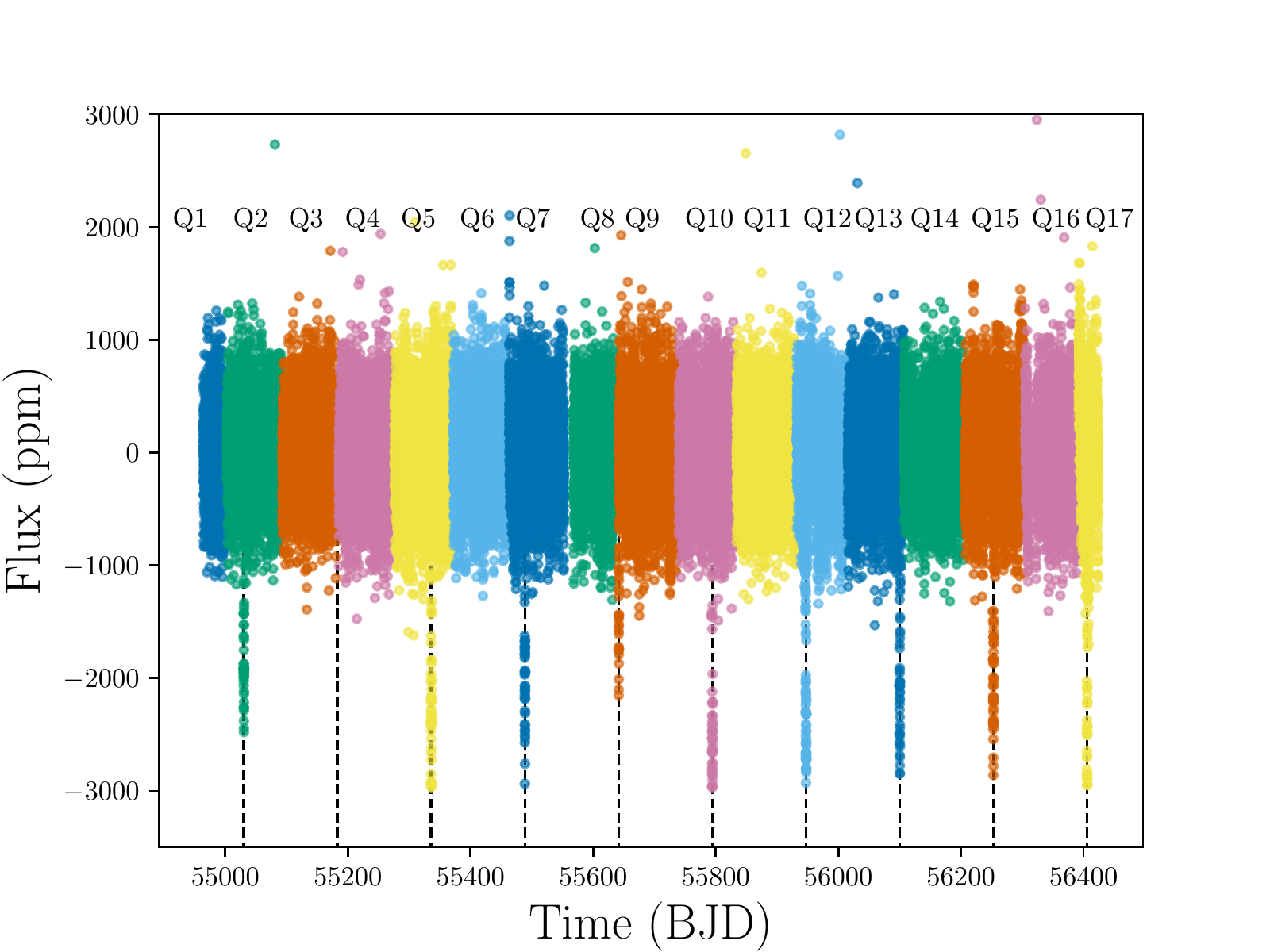}}
     
\caption{The left panel shows the lightcurve phase-folded on the orbital period of 152.8 days determined from the BLS search. To aid viewing the eclipse we have limited the plot in phase to $\pm0.2$. A smoothed version of the phase-folded data is given by the red dashed line and near the eclipse the additional variability caused by tidal interactions can be seen between a phase of $\pm0.05$. The right panel shows the detrended lightcurve of KOI-3890, showing clear eclipses, where the different colours denote different quarters of data as indicated. The vertical black dashed lines indicate the positions of the eclipses, and their expected locations in the event of there being a gap. The second eclipse that should have been observed falls in the gap between Q3 (orange) and Q4 (pink). The eclipse at the start of Q9 (orange) is only partially observed. }
\end{figure*}

\section{Observations}\label{sec:obs}

KOI-3890\footnote{KOI - \emph{Kepler} object of interest.} (KIC 8564976) was initially identified as a possible evolved-host planetary system based on \Kepler data using the NASA Exoplanet Archive \citep{ExArc2013}. When the transit depth (from the archive) was combined with the stellar radius, the resulting radius for the secondary clearly indicated the object is a stellar companion. The potential binary star nature of this system was first suggested in \cite{2015Lilobox}, who collected 22 radial velocities using the CAFE spectrograph \citep{2013Aceituno}. While the phase coverage in \cite{2015Lilobox} is limited, the authors place constraints on the minimum eccentricity $e\geq0.33$ and  minimum radial velocity amplitude $K\geq2.5\;\textrm{kms}^{-1}$, which is equivalent to a required minimum companion mass of $M>0.0097\pm0.0014\Ms$. This minimum mass is equivalent to $M>10.2\Mj$, and whilst this theoretically allows for a planet-mass object ($M\lesssim 13\Mj$; \citealt{RevModPhys.73.719}) this is more likely to be a binary star system. 

KOI-3890 was observed near-continuously for the 4 year duration of the \Kepler mission in long cadence (${\sim}30$ minutes) mode.  \Kepler detected a transit-like feature at a period of 152.8 days, and the system was flagged as a KOI. The long cadence (${\sim}30$ min) \emph{Kepler} lightcurve was downloaded from the Mikulski Archive for Space Telescopes (MAST)\footnote{https://archive.stsci.edu/index.html} and detrended with a moving median filter with a width of 20 days to ensure the eclipses were unaffected. The lightcurve was also clipped at the 4$\sigma$ level about the moving median. 

\subsection{Eclipses}
A box least-square (BLS) period search algorithm \citep{2002Kovacs} was used to detect the eclipses. The detrended lightcurve is shown in Fig~\ref{fig:3890tran}, whilst the lightcurve folded on the orbital period obtained from the BLS (152.8 days) is shown in Fig~\ref{fig:3890phase}. In Figure~\ref{fig:3890tran}, two of the ten potentially visible eclipses during the \Kepler mission are shown to be missing or only partially observed due to falling in gaps between quarters. If the lightcurve is folded on the period found by the BLS algorithm, it is clear that there is additional out of eclipse variability near to the phase of eclipse. The variability is shown in Fig~\ref{fig:3890phase} and is consistent with tidal interactions, as discussed in more detail in Sec~\ref{sec:3890model}. No secondary eclipses are detected in the data due to the geometry of the orbit (as explained in Sec~\ref{sec:diss3890}), as the impact parameter of the secondary eclipse is much greater than one.

\subsection{Spectroscopy}\label{sec:radvel}

In addition to the space-based \emph{Kepler} observations, we acquired spectroscopic data using the Tillinghast Reflector Echelle Spectrograph (TRES)  ($R=44,000$, $\lambda=3900-9100\AA$; \citealt{TRES}) on the 1.5-metre Tillinghast telescope at the Fred L. Whipple Observatory to measure radial velocities for the system. The TRES spectra were reduced and analysed using the techniques outlined in \cite{2010ApJ...720.1118B}. Multi-order velocities were derived by cross-correlating each spectrum, order by order, against the strongest observed spectrum (which had a signal-to-noise per resolution element of 29). Estimates of \teff, \logg, \mh were obtained using the Spectral Parameter Classification (SPC) technique described in \citealt{2012Natur.486..375B}, whereby an observed spectrum is cross-correlated against a grid of synthetic spectra based on \cite{1992IAUS..149..225K} model atmospheres. The asteroseismic value of \logg was used as a prior (such that the \logg was fixed to the asteroseismic value), and \mh (the relative metal abundance) was assumed to be equivalent to the metallicity \feh, the values are given Table~\ref{tab:3890params}. However, a reliable estimate of $\mathrm{v}\sin i$ was not able to be made for this system. In total, 10 radial velocity measurements were taken over 192 days, derived from multi-order fitting to spectral templates. The data taken are given in the Table~\ref{tab:3890_rv}.

In addition to the TRES observations, \cite{2015Lilobox} collected 22 radial velocities using the CAFE spectrograph \citep{2013Aceituno}, a high resolution spectrograph ($R=59000-67000$) in the optical range ($4000-9000\AA$) situated on the 2.2m telescope at Calar Alto Observatory. Although the phase coverage of the CAFE observations are rather limited they were also incorporated during the fitting of a model to the data since they occupy a region in phase that is not well covered by the TRES observations. The addition of the CAFE data trebled the available radial velocity measurements, giving a total of 32 RV points (22 from CAFE and 10 from TRES).

\section{Asteroseismic Analysis}\label{sec:astero}
As a red-giant star, KOI-3890A exhibits solar-like oscillations, driven by the turbulent convection in the near surface layers. We can use these oscillations to constrain the internal and global properties of the star. In this work we used the so-called ``global'' asteroseismic parameters: \numax, the frequency of maximum oscillation power; \delnu, the average large frequency spacing between modes of the same angular degree $\ell$ and subsequent radial order $n$; and $\Delta\Pi$ the period spacing. These parameters were extracted from the frequency power spectrum using the method described below (see also \citealt{Kallinger2014,2015A&A...578A..56K,2016Lund_hyades} for additional details). For the extraction of the asteroseismic parameters we used the frequency power spectrum of the detrended lightcurve (see Section~\ref{sec:obs}) constructed using the Lomb-Scargle periodogram \citep{1976Ap&SS..39..447L, 1982ApJ...263..835S} as provided by the python package \texttt{gatpsy} \citep{2015ApJ...812...18V, jake_vanderplas_2015_14833}. Due to there only being a few eclipses in the lightcurve, they were left in the timeseries. This did not affect the asteroseismic analysis due to the fact that the signal of the eclipses in the power spectrum is at a lower frequency ($<10\mu$Hz) than the asteroseismic signal ($\sim100\mu$Hz).

\subsection{Global asteroseismic parameters}\label{sec:fit}

\begin{figure}
\includegraphics[width=\columnwidth]{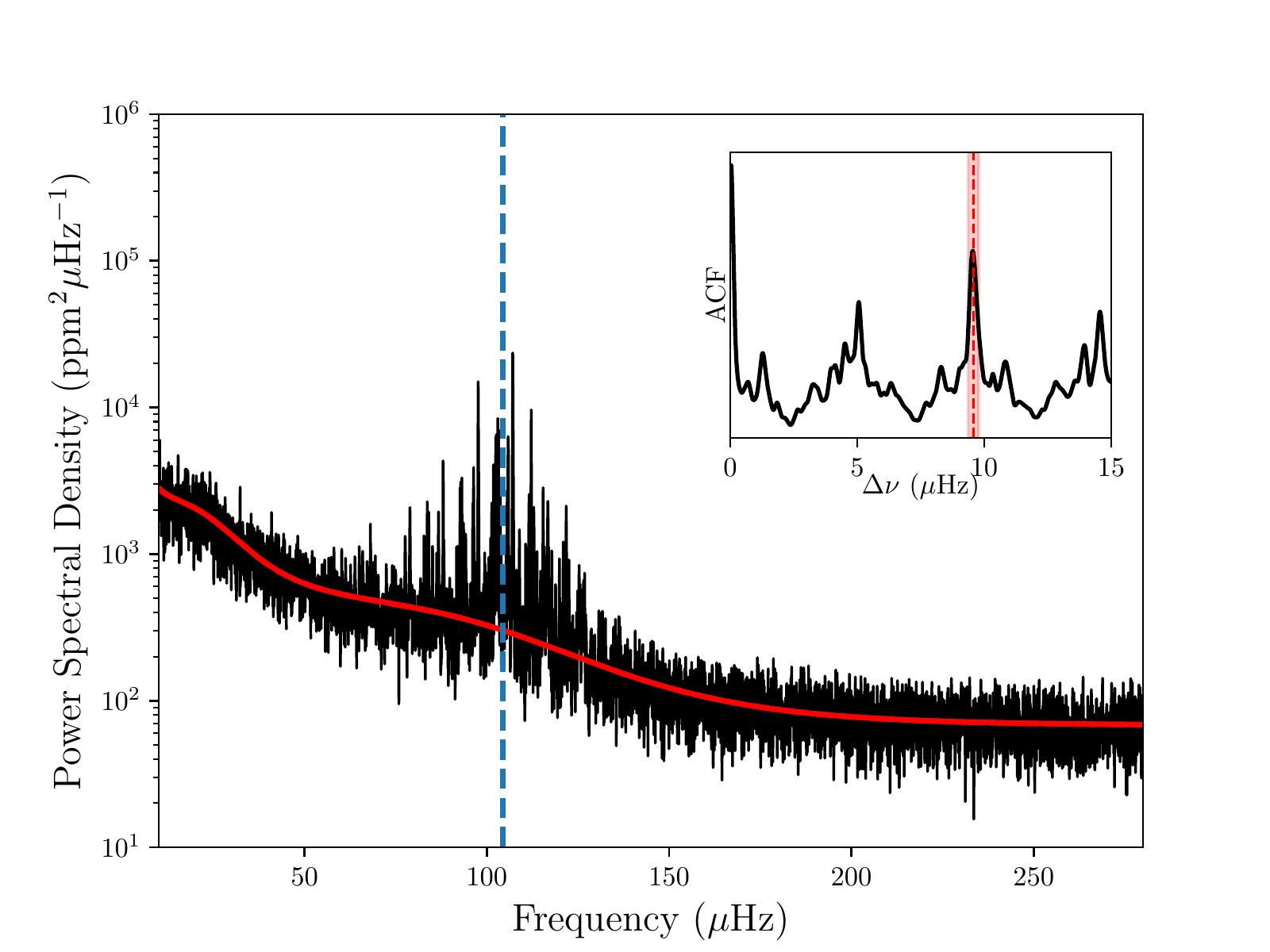}
\caption{The power spectrum of KOI-3890 is shown in black alongside the fit to the background (excluding the Gaussian component describing the power excess) in red. \numax is shown by the blue dashed line. In the inset we present the ACF showing a clear peak around \delnu with the red shaded region showing $1\sigma$ errorbars.}
\label{fig:psd}
\end{figure}

The first asteroseismic parameter we extract is \numax, which is determined by fitting a model to the background of the power spectrum. We adopted model H of \cite{Kallinger2014}, where the granulation background is modelled as two zero-frequency-centered ``super-Lorentzians'', where the Lorentzian is raised to an exponent that is a free parameter rather than the standard exponent of 2 or 4 (see \citealt{Kallinger2014}), along with one additional ``super-Lorentzian'' at very low frequency to account for any systematic effects. Also included in the background fit is a Gaussian component to account for the power excess due to the stellar oscillations centred on the frequency of maximum power, \numax. The fit was performed using MCMC and the python package \texttt{emcee} \citep{emcee}, the final fit parameters were taken as the median of the posterior distributions and the uncertainties given by the 68.3\% highest posterior density. The final fit to the background is shown in Fig \ref{fig:psd}. The inset in the figure shows the autocorrelation function of the power spectrum (ACF), used to identify the average large frequency separation \delnu, following \cite{2015A&A...578A..56K}. A Lorentzian was fitted to the ACF to determine $\Delta\nu$ (the main peak) and a conservative uncertainty was adopted as the half width at half maximum of the fitted Lorentzian to account for any addition deviations from the asymptotic pattern.

\begin{figure*}
\includegraphics[width=1.5\columnwidth]{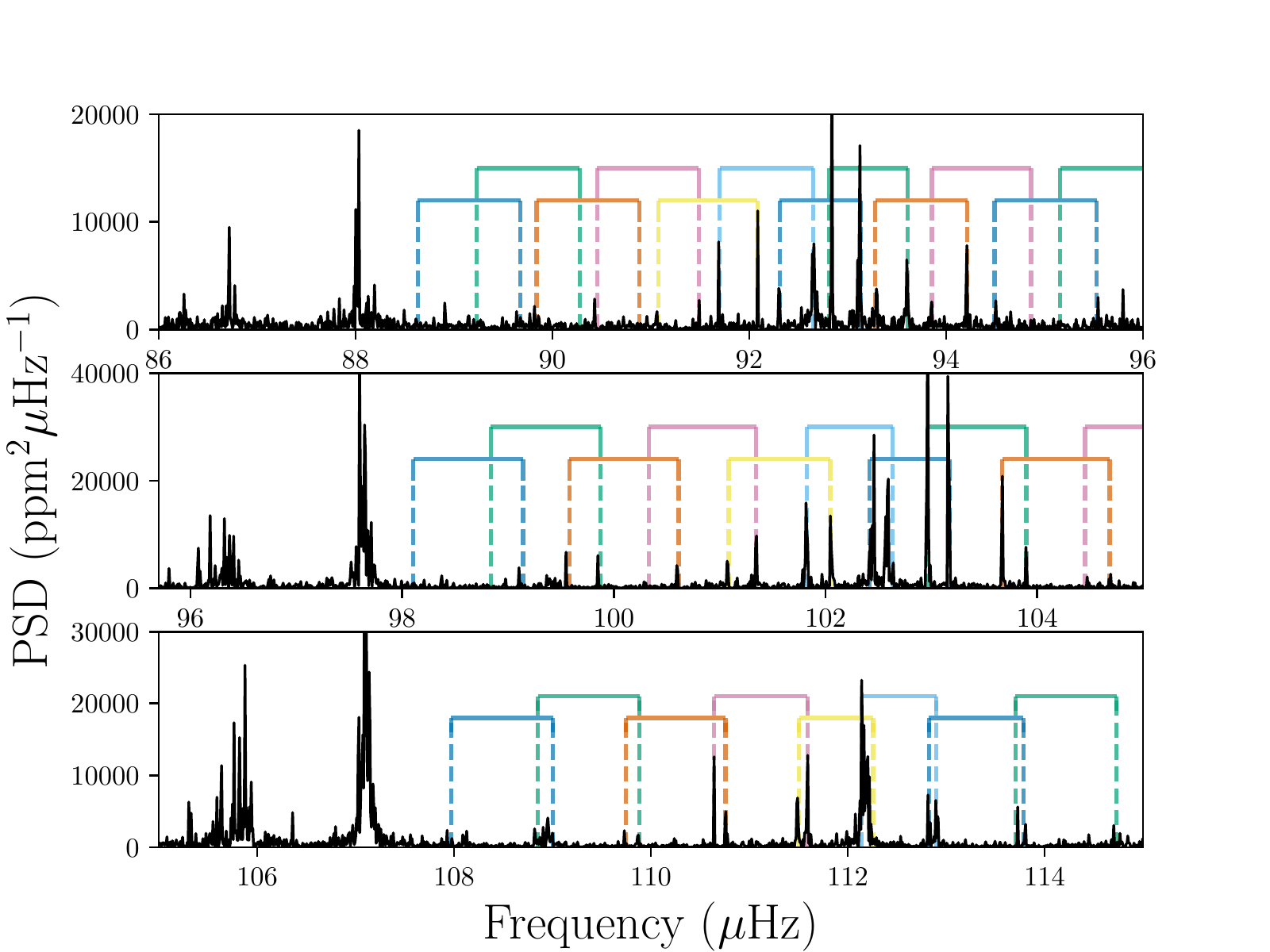}
\caption{A section of the KOI-3890 power spectrum showing the three radial orders used in the determination of the inclination angle. Each mixed mode is denoted by a different colour per radial order. No central $m=0$ components are seen indicating that the inclination angle is close to 90 degrees.}
\label{fig:psd_id}
\end{figure*}

In addition to \numax and \delnu, the period spacing (\dpi) of the $\ell=1$ mixed modes was also used to recover the stellar parameters. The mixed mode frequencies were extracted by extending the method described in \cite{2016AN....337..774D}, where we fit the $\ell=1$ mixed modes in the three radial orders around \numax at the same time (rather than using the properties of the mode components we already extracted). The priors on the parameters were all taken to be uniform with the exception of the inclination angle which was taken to be isotropic ($p(i)\propto\sin i$). The mode identification of the star given this model is shown in Fig~\ref{fig:psd_id}. The period spacing was then extracted from these frequencies using the method described in \cite{2018A&A...610A..80H}. \cite{Vrard2016} derived a period spacing of $\Delta\Pi_{1}=75.9\pm0.6$s for this star as part of their ensemble work by taking the power spectrum of the stretched power spectrum. We obtain a slightly higher period spacing than \cite{Vrard2016}, though still consistent and both show that the star is ascending the red-giant branch. To account for the difference in the methods, we adopt the conservative uncertainty from \cite{Vrard2016} on our period spacing value. The asteroseismic and spectroscopic parameters are summarised in Table \ref{tab:3890params}.

\begin{table}
	\centering
	\caption{Global asteroseismic and spectroscopic parameters for the red giant KOI-3890A.}
	\label{tab:3890params}
	\begin{tabular}{ll} 
		\hline
		\multicolumn{2}{c}{KOI-3890A}\\
		\hline
		\tabgap
		\numax &	104.3$\pm$0.3 $\mu$Hz\\
		\tabgap
		\delnu  &	9.57$\pm$0.21 $\mu$Hz\\
		\tabgap
		$\Delta\Pi$  & 77.6$\pm$0.6 s\\
		\tabgap
		\teff  & 4726$\pm$52 K\\
		\tabgap
		\feh  & -0.13$\pm$0.1 dex\\
		\tabgap
		$\mathrm{log}g_{\mathrm{astero}}$ & 2.92$\pm$0.03 dex\\
		\tabgap
		$\delta\nu_{\mathrm{rot, core}}$  & 0.533$\pm$0.003 $\mu$Hz\\
		\tabgap
		$i$  & $87.3^{+2.7}_{-1.1}$ degrees\\
		\tabgap		
		\hline
	\end{tabular}
\end{table}

\subsection{Rotational splittings and stellar inclination angle}\label{sec:inclination}

In addition to the global asteroseismic parameters, due to the high signal-to-noise ratio of the oscillations the rotational splittings and inclination angle can also be extracted for this star. The derivation of the stellar inclination angle (the angle between the rotation axis of the star and our line of sight) using asteroseismology adopts the formalism derived in \cite{Gizon2003}. Since stars rotate, modes of the same $n$ and $l$ and different $m$ are not degenerate with one another. In the case of non-radial ($l>0$) modes the relative amplitude of each component provides information about the inclination angle of the star, whilst the frequency difference provides the rotational splitting. For red giants, the inclination angle and rotational splitting can be measured using $l=1$ mixed modes. The inclination angle posterior probability distribution was extracted from the fit to the oscillation frequencies detailed in Section~\ref{sec:fit}, giving a value of $i(^{\circ})=87.3^{+2.7}_{-1.1}$. An estimate of the core rotation can be attained from the rotational splitting of the modes (following the formulation of \citealt{2018arXiv180708301M}), as given in Table~\ref{tab:3890params}. This is broadly consistent with the value obtained by \cite{2018A&A...616A..24G} of $\delta\nu_{\mathrm{rot, core}}=0.520\pm0.002 \mu$Hz, allowing for differences in formulation and method.  

The frequency spacing between consecutive mixed $\ell=1$ modes in the power spectrum of the red-giant star is comparable to the rotational splitting, which complicates the mode identification. In order to check that we have the correct mode identification, we investigate the rotational splitting inferred from each radial order, which should be consistent throughout. If overlapping modes are not assumed then the rotational splittings decrease significantly as a function of radial order, which is not physical. Therefore the interpretation of overlapping modes is needed and the first guesses in the peak-bagging have been updated accordingly.

Given the eclipsing nature of the system, it is also possible to determine the obliquity ($\psi$) of the system. The obliquity of the system is related to the inclination angle through the following equation \citep{2009ApJ...696.1230F,2014MortonWinn}

\begin{equation}
\sin\psi\cos\phi = \sin i\cos\lambda\cos i_{\mathrm{p}} - \cos i\sin i_{\mathrm{p}},
\label{eqn: ob}
\end{equation}
where $\lambda$ is the sky-projected obliquity that can be determined from the Rossiter-MacLaughlin effect and $\phi$ is the azimuthal angle of the system. Since the system is eclipsing, $\sin i_{\mathrm{p}}\approx 1$, and so $\cos i_{\mathrm{p}}\approx 0$. Following \cite{2014MortonWinn}, Eq~\ref{eqn: ob} can be reduced to
\begin{equation}
\sin\psi\cos\phi\approx\cos i.
\end{equation}
Since we cannot distinguish between the angles $i$ and $\pi-i$ when inferring the inclination angle the negative sign in Eqn.~\ref{eqn: ob} can be ignored.

\begin{figure*}
     \centering
     \subfloat[][]{\includegraphics[width=\columnwidth]{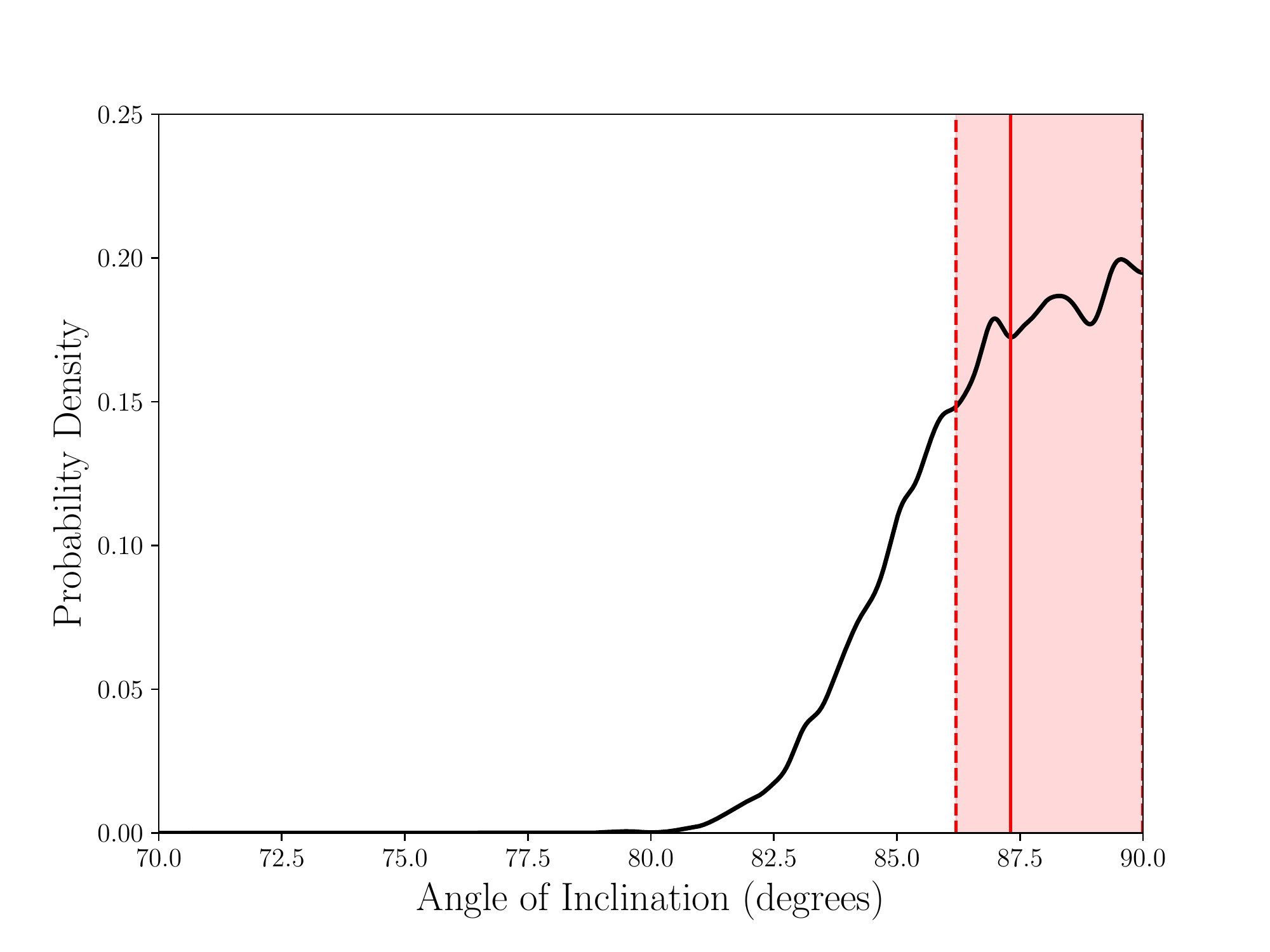}}
     \hfill
     \subfloat[][]{\includegraphics[width=\columnwidth]{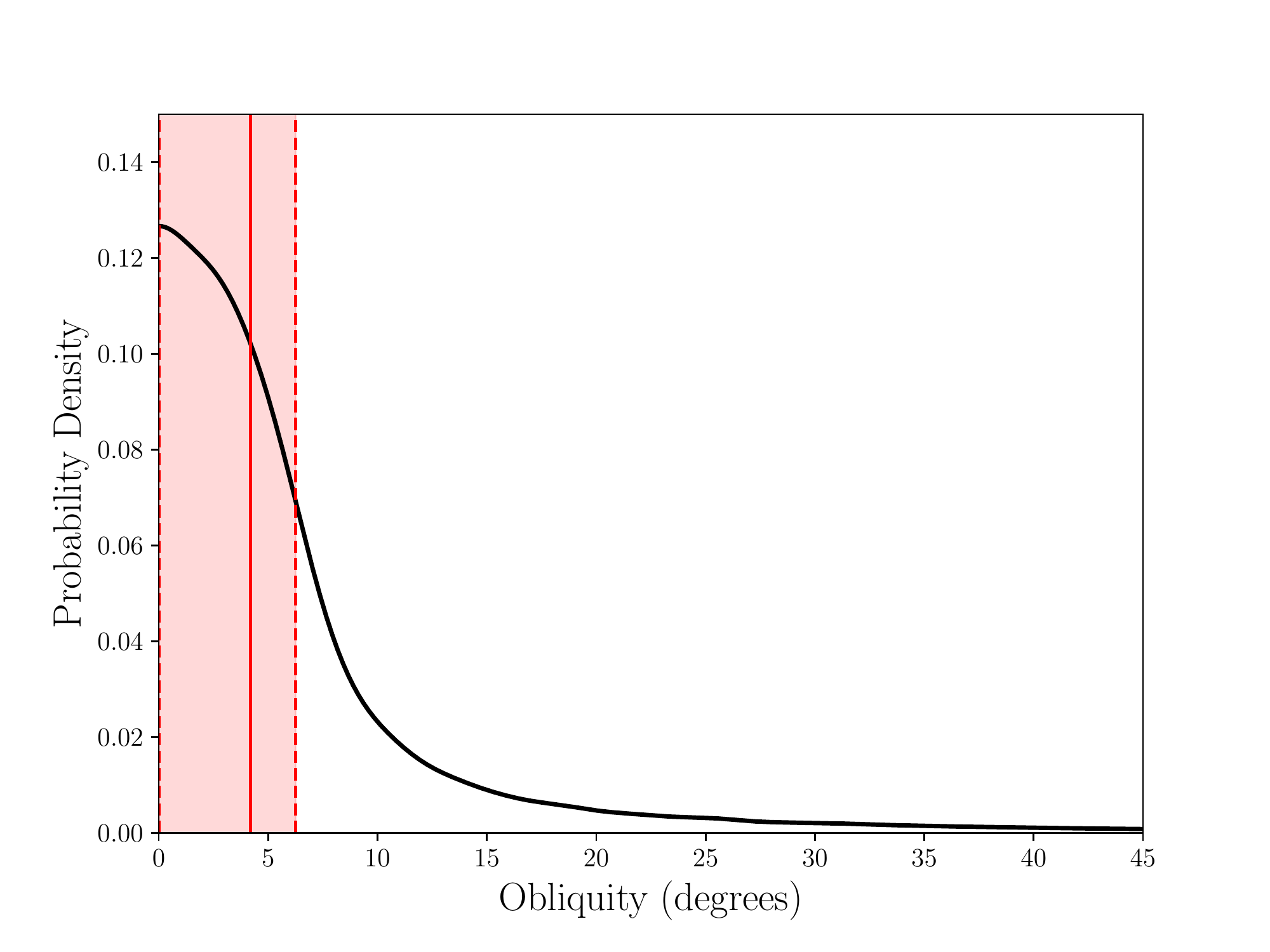}}
     \caption{Kernel density estimates (KDE) of the inclination angle posterior probability distribution (left panel) and obliquity (right panel). The median value is shown by the solid red line, whilst the 68.3\% highest posterior density (HPD) credible interval is encompassed by the shaded region and dashed lines.}
     \label{fig:ob}
\end{figure*}

The azimuthal angle varies between $-\pi$ and $\pi$, where $\pi$ is defined as being along the line of sight \citep{2009ApJ...696.1230F}. The obliquity of the system can be estimated from a Monte-Carlo approach using the inclination angle posterior PDF extracted from the fitting. We also assume that the azimuthal angle is distributed uniformly between $-\pi$ and $\pi$ since from asteroseismology we cannot make any inferences about it. The obliquity of the system was found to be $\psi(^{\circ})=4.2^{+2.1}_{-4.2}$, the posterior of which is shown in Fig~\ref{fig:ob}. This is consistent with alignment, meaning that the plane of the orbit of the companion star is perpendicular to the stellar rotation axis of the primary (e.g. the Earth has an obliquity of ${\sim}23^{\circ}$ to its orbital plane).

\section{Stellar Properties}\label{sec:param}
The global asteroseismic (including the observed $\Delta\nu$ and period spacing, $\Delta\Pi_{1}$), and spectroscopic parameters are used as inputs to stellar models to recover the estimated stellar properties of each component in the binary system. The Bayesian code \textsc{PARAM}\footnote{\url{http://stev.oapd.inaf.it/cgi-bin/param}} \citep{2006dasilva, 2014Rodrigues,2017Rod} was used to recover the fundamental stellar properties, using the MESA isochrones from \citealt{2017Rod} (see section 2 of \citealt{2017Rod} for full details regarding the input physics used in the creation of the models). In PARAM, \delnu was calculated from theoretical radial mode frequencies using a weighted linear fit, where the weights are inversely proportional to the distance between the given frequency and \numax. This is to ensure that the computed average \delnu is a close as possible to the observational value (see \citealt{2017Rod} for more information).

Using the above physics, we obtain the fundamental stellar properties of the red giant, given in Table \ref{tab:stellar}. These will enable the characterisation of the companion star from the orbital parameters derived using the radial velocity and photometric observations.

\begin{table}
	\centering
	\caption{Fundamental stellar properties from \textsc{PARAM}, quoted uncertainties are the 68\% credible interval.}
	\label{tab:stellar}
	\begin{tabular}{ll} 
		\hline
		\multicolumn{2}{c}{KOI-3890A}\\
		\hline
		\tabgap
		$M_{\star}$  &	$1.04\pm0.06$ $\Ms$\\
		\tabgap
		$R_{\star}$  &	$5.8\pm0.2$ $\textrm{R}_{\odot}$\\
		\tabgap
		Age  & $9.1^{+2.4}_{-1.7}$ Gyr\\
		\tabgap
		\hline
	\end{tabular}
\end{table}

\section{Lightcurve and Radial Velocity Modelling}\label{sec:3890model}
The transit and radial velocity data for KOI-3890 are modelled simultaneously, using MCMC\footnote{Using Python package \textsc{emcee}  \citep{emcee}} to sample the parameter space. Only phases in the folded lightcurve of $-0.2<\phi<0.2$ were retained when fitting the full unfolded transit photometry as the features caused by the tidal interactions occur near the eclipse. The eclipses with incomplete phase coverage and surrounding data were also removed from the lightcurve.

The radial velocity data were modelled as,
\begin{equation}
V(t)=\gamma + K\left[\cos(f(t)+\omega) + e\cos{\omega}\right],
\label{eqn:radvel}
\end{equation}
where $f(t)$ is the true anomaly, $e$ the orbital eccentricity, $\omega$ the argument of periastron, $K$ the radial velocity semi-amplitude and $\gamma$ is the zero point offset of the Doppler velocities. The initial guess for the period was taken from the BLS detection routine.

The Python package \textsc{Batman} \citep{Batman2015} was used to model the eclipse, based upon the transit model of \cite{2002MandelAgol}. We assume that the secondary is not a self-luminous body and we discuss this in more detail in Section~\ref{sec:diss3890}. Limb darkening parameters were taken from \cite{2010Sing}, and were fixed in the transit model, using a quadratic formulation for the limb darkening law. Additional parameters $\sigma_{\text{RV,CAFE}}$ and $\sigma_{\text{RV,TRES}}$ are included to account for any additional variance not accounted for in the original observational uncertainties for the respective radial velocity datasets. Since the \Kepler observations are integrated over 30 minutes, the transit model is supersampled by a factor 5. If this is not done, the transit duration can be underestimated \citep{Batman2015}.  

The initial fit to the TRES radial velocity data indicated the orbit was highly eccentric, with the eclipses being observed near periastron. 
For the final fitting, the TRES data were combined with the CAFE radial velocities from \cite{2015Lilobox}, with an additional parameter included in the fit to account for zero-point offset between the data sets.

\begin{table}
	\centering
	\caption{Radial velocity data taken with TRES used in this work.}
	\label{tab:3890_rv}
	\begin{tabular}{lll} 
	\hline
	Time (BJD) & Velocity ($\mathrm{ms}^{-1}$) & Uncertainty ($\mathrm{ms}^{-1}$)\\
	\hline
	2457171.774856 &  -1597.01  &    16.79\\
	2457210.763886 &      0.00  &    41.42\\
	2457237.798983 &   4006.68  &    22.57\\
	2457289.755597 &  10216.62  &    46.41\\
	2457296.785214 &  11074.15  &    47.32\\
	2457304.778182 &  11853.78  &    41.42\\
	2457319.633079 &   7545.42  &    32.03\\
	2457349.598870 &  -2987.42  &    30.57\\
	2457640.650105 &  -7313.28  &    41.35\\
	2457932.777215 &   4686.70  &    53.45\\
	\hline
	\end{tabular}
\end{table}

\subsection{Tidal distortion of primary}
The phase folded lightcurve around the time of mid-transit is shown in Figure \ref{fig:3890phase}, where the additional flux variation just outside the transit can be seen. As discussed previously, the flux contribution from the secondary component is negligible. It can therefore be assumed that the flux variation is originating from the primary (giant) star. 

This additional flux variation appears to be indicative of a heartbeat star, a type of tidally induced variation \citep{2011Welsh,2012Thompson,2014Beck,2016Shporer,2018arXiv180305917P}. For circular orbits ($e=0$), this tidal effect is constant, raising a bulge on the primary star. In the case of this highly-eccentric system it produces ellipsoidal variation. For eccentric orbits, the term ``heartbeat'' reflects the passing visual similarity of the tidally induced variation to an echocardiogram.

The heartbeat is induced when the tidal distortions are largest near periastron. Whilst this system has a fairly long period, its eccentricity and the evolved nature of the primary star means that, at periastron passage, the separation between the stars is only a few times the primary radius, and so the secondary is able to raise a tidal bulge on the primary.

The heartbeat flux modulation was modelled following the prescription given in \cite{1995Kumar}, adjusting for the fact that we are modelling it as a function of the true anomaly,
\begin{equation}
\delta F=S\cdot \frac{1-3\sin^{2}i_{\mathrm{p}}\cos^{2}\left[f(t)-\omega\right]}{[R(t)/a]^{3}}.
\label{eq:tide}
\end{equation}
In Eq~\ref{eq:tide}, $S$ is the amplitude of the flux variation, $i_{\mathrm{p}}$ the system inclination angle (known to be $i_{\mathrm{p}}\sim 90^{\circ}$ since the eclipse is observed), $f(t)$ is the true anomaly, $\omega$ the argument of periastron, and $R(t)/a$ the distance between the two stars as a fraction of the semi-major axis $a$, as a function of time. $R(t)/a$ can also be expressed as (Eq 1 \citealt{Winn2010}),
\begin{equation}
\frac{R(t)}{a}=\frac{1-e^2}{1+e\cos f(t)}.
\label{eq:rt_a}
\end{equation}

The addition of the flux modulation into the lightcurve modelling through the inclusion of Eqs \ref{eq:tide} and \ref{eq:rt_a} introduces only one new parameter, the amplitude of the heartbeat modulation $S$. All other parameters are already included within the transit and radial velocity models. 

The formulation for the heartbeat signal can produce a wide variety of possible lightcurve modulations, due to the possible orientations of the system in $i_{\mathrm{p}}$ and $\omega$ (see Figure 8 of \citealt{2012Thompson}). In the case of KOI-3890, the presence of eclipses indicates that the system is close to edge-on $(i_{\mathrm{p}}{\sim}90^{\circ})$, and so fitting the heartbeat modulation simultaneously with the transit model offers a tighter constraint on $i_{\mathrm{p}}$.

An additional prior can be applied to the stellar density due to the fact that the red giant primary shows solar-like oscillations. The average large frequency separation \delnu scales to good approximation with the square-root of mean stellar density \citep{1980Tassoul,1986ApJ...306L..37U}, i.e.,
\begin{equation}
\label{eq:delnu_host}
\frac{\Delta\nu}{\Delta\nu_{\odot}} \simeq \left(\frac{M}{\text{M}_{\odot}}\right)^{0.5}\left(\frac{R}{\text{R}_{\odot}}\right)^{-1.5}\simeq\sqrt{\frac{\rho}{\rho_{\odot}}}.
\end{equation}
The mean stellar density can also be estimated from the lightcurve using Eq 30 of \cite{Winn2010},
\begin{equation}
\rho_{\star}\approx\frac{3\pi}{GP^2}\left(\frac{a}{R_{\star}}\right).
\label{eq:tran_density}
\end{equation}
where $a/R_{\star}$ is taken from the transit fit.

During the fitting, the stellar density at each iteration was constructed using Eq \ref{eq:tran_density}, and the asteroseismic density (determined through PARAM) used as a Gaussian prior. The priors used during the simultaneous fit to the radial velocity and photometric data are given in Table \ref{tab:3890_priors}.

\section{Results}
\begin{figure}
\centering
\subfloat[][]{\includegraphics[width=\columnwidth]{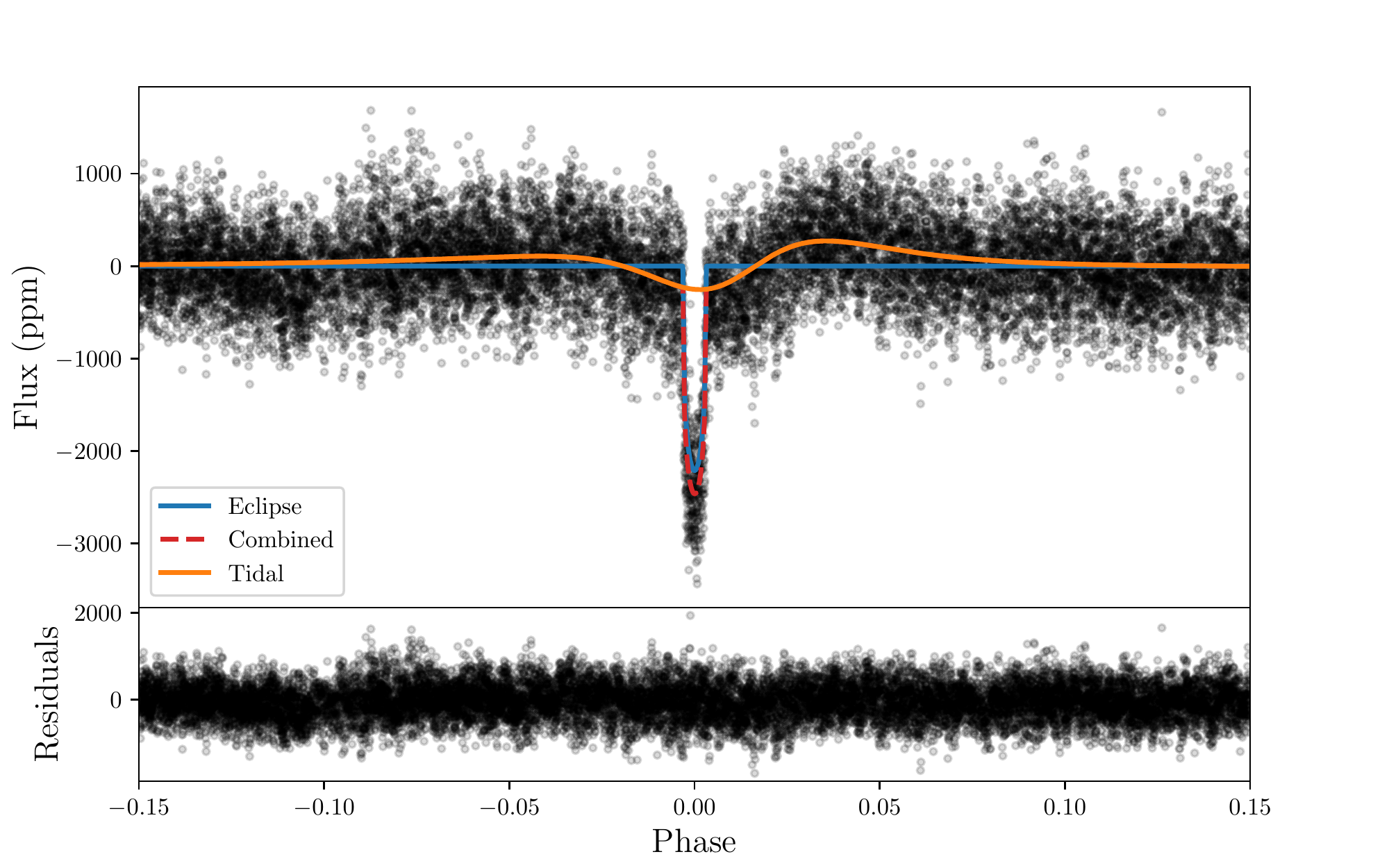}}
\hfill
\subfloat[][\label{fig:3890_final_rv}]{\includegraphics[width=\columnwidth]{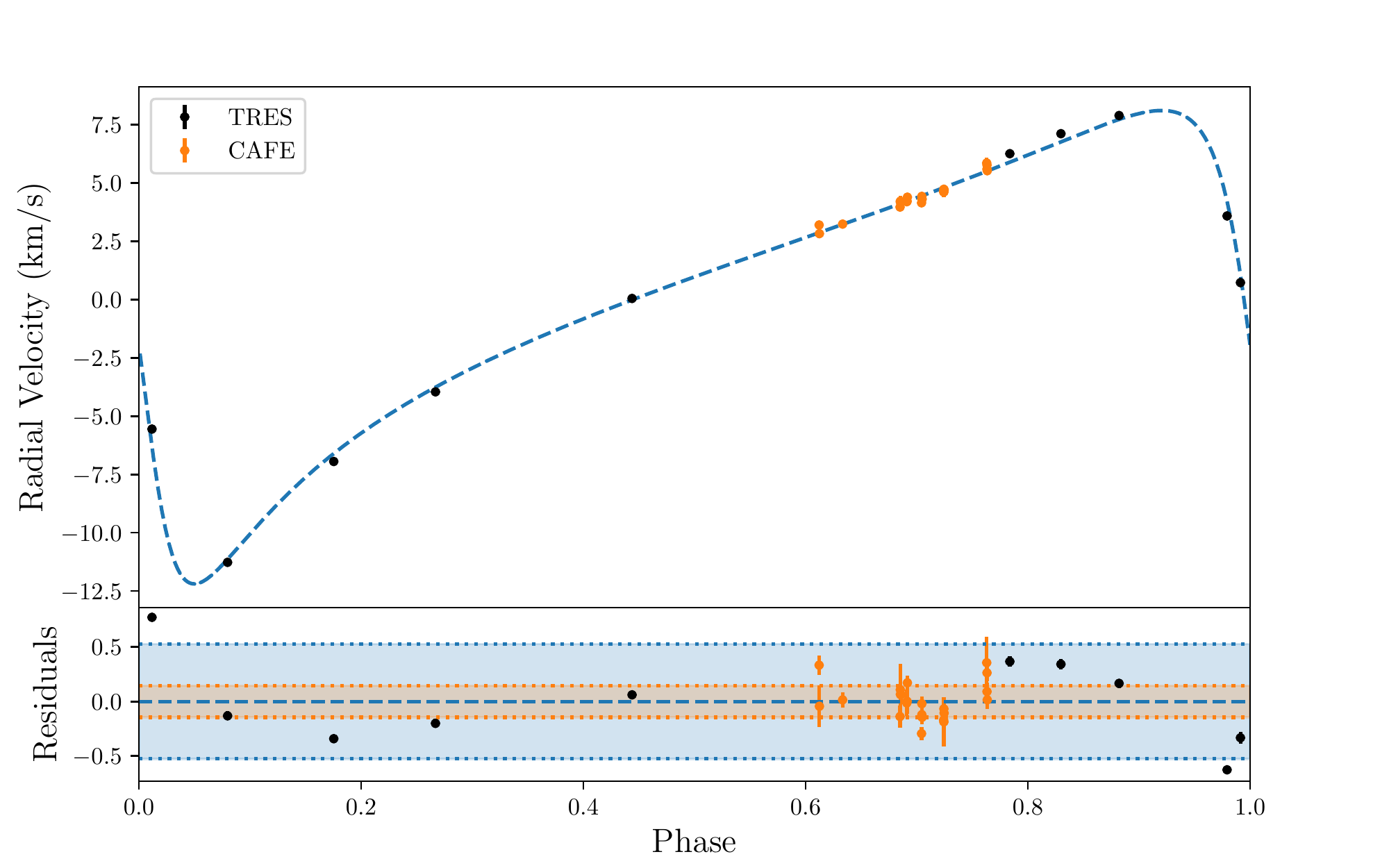}}
\caption{Phase folded final fit centred around time of mid transit including residual plots. In the upper panel, the lightcurve is in black, with the final model shown dashed in red near the eclipse. The two components of the lightcurve model are also shown, the transit in blue, and the tidal model in green. In the lower panel the TRES radial velocity data (black) and CAFE data (orange) are shown phase folded and also centred around mid transit, with the final model shown as dashed blue. The orange and blue shaded regions correspond to the addtion variance terms (see Section~\ref{sec:3890model}) for the CAFE and TRES data respectively.}
\label{fig:3890_final}
\end{figure}

 The results of the combined transit, tidal and radial velocity MCMC fit are listed in Table \ref{tab:3890fit}. Figure \ref{fig:3890_final} shows the final fit of the model to the data. As the upper panel of Fig \ref{fig:3890_final} shows, the tidal distortion of the primary would artificially boost the eclipse depth if not properly accounted for. This can be an important factor when determining the radius of the secondary since it will be overestimated if the tidal distortion is not included. In the fully convective regime\textemdash $M_{\star} \lesssim 0.33 \mathrm{M}_{\odot}$ \citep{2015ApJ...804...64M}\textemdash it is expected that the ratio of the mass and radius of an M-dwarf is approximately unity (with respect to the Sun, e.g. \citealt{2009A&A...505..205D}). Therefore not accounting for tidal effects would indicate a larger radius than otherwise expected and could lead to the incorrect conclusions.

\begin{table}
	\centering
	\caption{Model prior distributions for the simultaneous fit to the radial velocity and photometric data. Gaussian priors indicated by $\mathcal{N}$(mean,standard deviation), and uniform priors by $\mathcal{U}$(lower bound, upper bound). All logarithmic priors are in base 10. }
	\label{tab:3890_priors}
	\begin{tabular}{ll} 
	\hline
	Parameter & Prior\\
	\hline
	$P$ &	$\mathcal{U}(151,153)$ (days)\\
	$\log(R_{2}/R_{\star})$&$\mathcal{U}(-3,-1)$\\
	$\log(a/R_{\star})$ & $\mathcal{U}(0,2)$\\
	$T_{0}$ & $\mathcal{U}(55025,55035)$ (BJD)\\
	$b$ & $\mathcal{U}(0,1)$\\
	$e\cos{\omega}$ & $\mathcal{U}(-1,1)$\\
	$e\sin{\omega}$ & $\mathcal{U}(-1,1)$\\
	$\gamma_{\textrm{TRES}}$ & $\mathcal{U}(-100,100)$ (km/s)\\
	$\gamma_{\textrm{CAFE}}$ & $\mathcal{U}(-100,100)$ (km/s)\\
	$\log(K)$ & $\mathcal{U}(-1,2)$ (km/s)\\
	$\sigma_{\text{RV,TRES}}$ & $\mathcal{U}(0,10)$ (km/s)\\
	$\sigma_{\text{RV,CAFE}}$ & $\mathcal{U}(0,10)$ (km/s)\\
	$S$ & $\mathcal{U}(-100,100)$ (ppm) \\
	$\rho_{\star}$ & $\mathcal{N}(7.09,0.31)$ (kg/m$^{3}$)\\	
	\hline
	\end{tabular}
\end{table} 

\begin{table}
	\centering
	\caption{Model parameters median values from the simultaneous fit to the radial velocity and photometric data, and associated uncertainties taken as the 68.3\% credible interval.}
	\label{tab:3890fit}
	\begin{tabular}{ll} 
	\hline
	Parameter & Median Value\\
	\hline
	$P$ &	$152.826\pm0.0002$ (days)\\
	$R_{2}/R_{\star}$& $0.0444\pm0.0002$\\
	$a/R_{\star}$ & $20.44\pm0.28$\\
	$T_{0}$ & $55030.411\pm0.001$ (BJD)\\
	$i_{\mathrm{p}}$ & $85.3\pm0.2$ (deg)\\
	$e$ & $0.645\pm0.001$\\
	$\omega$ & $108.7\pm0.02$ (deg)\\
	$\gamma_{\textrm{TRES}}$ & $3.96\pm0.20$ (km/s)\\
	$\gamma_{\textrm{CAFE}}$ & $-30.6\pm0.13$ (km/s)\\
	$K$ & $10.1\pm0.3$ (km/s)\\
	$\sigma_{\text{RV,TRES}}$ & $0.53\pm0.16$ (km/s)\\
	$\sigma_{\text{RV,CAFE}}$ & $0.14\pm0.03$ (km/s)\\
	$S$ & $-12.0\pm0.1$ (ppm) \\
	\hline
	\end{tabular}
\end{table} 

In order to calculate the secondary mass, we make use of the following expansion for $K$ (e.g. \citealt{2010arXiv1001.2010W})
\begin{equation}
    K = \frac{1}{\sqrt{1-e^{2}}}\left(\frac{P}{2\pi G}\right)^{-1/3}\frac{M_{2}\sin i_{\mathrm{p}}}{(M_{1} + M_{2})^{2/3}},
\label{Eq:K_exp}
\end{equation}
where $M_{1}$ is the mass of the red-giant primary and $M_{2}$ is the mass of the M-dwarf companion. The secondary mass was calculated by numerically solving Eq \ref{Eq:K_exp} with a Monte-Carlo method that draws samples from the posterior distributions of each parameter. The derived mass and radius of the secondary are given in Table \ref{tab:3890_second}.
\begin{table}
	\centering
	\caption{Derived mass and radius for the secondary from the fit and the asteroseismic mass and radius of the primary. Both mass and radius of the secondary are consistent with an M dwarf.}
	\label{tab:3890_second}
	\begin{tabular}{ll}
	\hline
	\multicolumn{2}{c}{KOI-3890B}\\ 
	\hline
	$M_{2}$ & $0.23\pm0.01$ \Ms \\	
	$R_{2}$ 
	& $0.256\pm0.007$ \Rs \\	
	\hline
	\end{tabular}
\end{table} 

\section{Discussion and conclusions}\label{sec:diss3890}

In this work we have presented the analysis of the red-giant/M-dwarf eclipsing binary system, KOI-3890. We have determined the stellar parameters of both the primary and secondary companion along with the eclipse parameters by including tidal distortions into the fitting process.

We have made an assumption during the analysis that it is valid to fit the eclipses as transits since we do not have spectroscopic information regarding the properties of the M-dwarf companion \textemdash to treat the M-dwarf as though it is not a self-luminous body. This assumption can be justified  by considering the relative flux contribution of the two components in the system. Using the derived radius of the secondary and assuming $T_{\mathrm{eff},2}\approx3000\textrm{K}$ (see \citealt{Kaltenegger2009}) we obtain a luminosity of $L_{2}\approx 0.005\textrm{L}_{\odot}$ compared to $L_{1}\approx 15\textrm{L}_{\odot}$, a factor of almost 3000 in luminosity.

There is some tension present in the combined fitting of the photometry and radial velocity data as can be seen by the large unaccounted variance parameter needed for the TRES data and by the systematic trends induced in the TRES radial velocity data (see Fig.~\ref{fig:3890_final_rv}). This is due to a difference in eccentricity obtained from the photometry and radial velocity. A fit to only the radial-velocity data yields $e=0.608\pm0.001$ as opposed to the larger value favoured by the photometry (as seen in the combined fit). This is because the eccentricity can also be constrained in the photometry through the heartbeat signal. We have much more data in the photometric dataset than the radial velocity dataset and so as a result the photometry dominates the fit leading to the larger eccentricity. \cite{2011Welsh} also noticed this type of discrepancy between the fit to the radial velocity data alone and a combined fit to both the photometry and radial velocities.

We put forward two possible explanations for the discrepancy in the eccentricities obtained from the radial velocity data and the photometry:
\begin{enumerate}
	\item The observed radial velocities may not correspond to the centre of mass of the tidally distorted star. The observed data corresponds to the value integrated across the distorted surface weighted by the intensity for each surface element, which could distort the apparent eccentricity inferred from the velocities.
	\item The detrending of the data may affect the amplitude and shape of the heartbeat signal in subtle ways that cannot easily be identified, yet still lead to a different eccentricity inferred overall. 
\end{enumerate}

Since the heartbeat signal offers a great ability to constrain the eccentricity, any additional physical features not accounted for could potentially bias the eccentricity upwards because the amplitude of the heartbeat scales with the eccentricity. This does not change the interpretation of the system, but it means that we can only reliably say that $0.6 \leq e \leq 0.65$ (this distribution is used in all simulations requiring the eccentricity), where the lower limit is set by the radial velocity data and the upper limit by the photometry.

The mass of the companion                                                                                                                                                                                                                                                                                                                                                                                                                                                                                                                                                                                                                                                                                                                                                                                                                                                                                                                                                                                                                                                                                                                                                                                                                                                                                                                                                                                                                                                                                                                                                                                                                                                                                                                                                                                                                                                                                                                                                                                                                                                                                                                                                                                                                                                                                                                                                                                                                                                                                                                                                                                                                                                                                                                                                                                                                                                                                                                                                                                                                                                                                                                                                                                                                                                                                                                                                                                                                                                                                                                                                                            has been computed using the eccentricity from the combined fit ($M_{2}=0.23\pm0.01\textrm{M}_{\odot}$), the radial velocity data only ($M_{2}=0.23\pm0.01\textrm{M}_{\odot}$) and the eccentricity range we can infer $0.6 \leq e \leq 0.65$ ($M_{2}=0.23\pm0.01\textrm{M}_{\odot}$). In all of these cases the mass of the M-dwarf companion is consistent and the uncertainty in the eccentricity measurement does not affect the mass measurement. This is due to the dominant uncertainty coming from the radial-velocity semi-amplitude $K$ which is consistent between the combined and radial-velocity only fit. Hence the eccentricity uncertainty does not affect the interpretation of the system and so the mass of the companion given is for the eccentricity range $0.6 \leq e \leq 0.65$. However, these systematics could also propagate through to the eclipse depth, thereby affecting the radius of the secondary. 

The presence of tidally-induced distortions of the primary star is clearly seen in the lightcurve of KOI-3890, and these ``heartbeat'' events have been included in the eclipse fitting. We do not observe any tidally-induced oscillations in our data, but this have been observed in other systems \citep[e.g][]{2017ApJ...834...59G, 2018MNRAS.473.5165H}. 

As discussed earlier in Section~\ref{sec:obs}, the lack of secondary eclipses (the secondary passing behind the primary from the observer's perspective) is due to the inclination and eccentric nature of the system. The impact parameter $b$ for transiting or occulting objects is given in \citet[Eq 7,8]{Winn2010},
\begin{equation}
b=\frac{a \cos i_{\mathrm{p}}}{R_{\star}}\left(\frac{(1- e^{2})}{1\pm e\sin{\omega}}\right),
\label{eq:ecc_b}
\end{equation}
where eclipses are (+) and occultations (-) in the denominator. For KOI-3890, the eclipse impact parameter $b_{\textrm{tran}}=0.64\pm0.04$, however for the secondary eclipse $b_{\textrm{sec}}=2.5\pm0.1$. As a result, the M-dwarf does not pass behind the primary during its orbit, and so no secondary is observed, since $b\leq 1$ is required for an eclipse or secondary eclipse to occur.

It has been shown that some M-dwarfs in binary systems (where the parameters of both components can be precisely determined) show evidence of radius inflation compared to stellar models (e.g. \citealt{2012MNRAS.426.1507B, 2005ApJ...631.1120L}; and references therein). Magnetic activity inhibiting convection and causing the star to ``puff-up'' is a possible explanation to describe such behaviour \cite[][e.g.]{2007A&A...472L..17C, 0004-637X-765-2-126} . \cite{0004-637X-757-1-42} favour a so-called ``hybrid interpretation'', whereby they suggested that alongside magnetic-activity. Radius inflation could also be a result of unseen systematics in the determination of the radius. This could come about because almost all measurements of M-dwarfs come from close binaries where the components are tidally locked, active and rapidly rotating. However, \cite{2018AJ....155..225K} show that neither rotation nor binarity are responsible for the observed inflated radii. 

Since the majority of cases where accurate M-dwarf masses and radii are available generally come from the aforementioned eclipsing systems, there are only a small number since they must be close-by to be analysed with current instrumentation. \cite{2018MNRAS.tmp.2233P} showed that another useful regime for inferring M-dwarf properties is M-dwarf/white-dwarf eclipsing systems. They analysed 23 such systems, finding that around 75\% of their sample were inflated compared to theoretical models.

KOI-3890 provides a means to look at M-dwarf properties in a very different regime to the generally close-in, short-period M-dwarf/M-dwarf eclipsing systems  (e.g. \citealt{2009A&A...505..205D}). We do, however, note that this is a special case since it is in a position that is highly influenced by the red-giant primary (due to the close proximity at periastron). The general idea of using the red-giant/M-dwarf regime is still valid, since the application of asteroseismology to such eclipsing binary systems can help provide additional constraints to M-dwarf stellar models, especially in the fully-convective regime where the models typically struggle to reproduce the observations. The effect of systematics is something we have already touched on regarding the inclusion of the tidal distortion in the modelling of the eclipse. We can assess whether the M-dwarf companion in our system agrees with stellar models by comparing the derived mass and radius with the models from \cite{2015A&A...577A..42B} at a similar age. The closest stellar models in age to our system are those at 8 Gyr and 10 Gyr, however the mass and radius of the M-dwarf do not change significantly over this period and so we compare to the model at 8 Gyr. Whilst the radius for KOI-3890B is larger than the predicted value from the models, $R_{\star}/R_{\mathrm{model}} = 1.06\pm0.05$, it is consistent with the model at just over the $1\sigma$ level. The models were, however, computed for solar metallicity and since KOI-3890 is slightly metal-poor when compared to the Sun (if we assume the M-dwarf has the same metallicity as the red giant primary) this may have an additional effect.

Alongside providing good constraints on the mass and radius of the M-dwarf companion through asteroseismology, it is also possible to place constraints on the age of the system, of $9.1^{+2.4}_{-1.7}$ Gyr. This is difficult for the case of the M-dwarf/M-dwarf eclipsing systems as they are reliant on the low-mass stellar models which are known to suffer from inconsistencies in the fully-convective regime. In addition, the change in luminosity and effective temperature as a function of age is very small making it more difficult to constrain M-dwarf ages from stellar models. Whereas for KOI-3890, we can take advantage of the advances made in the modelling of red-giant stars using asteroseismic constraints to provide an age for the system. Even though the system is a single-lined spectroscopic binary, the asteroseismic constraints can help make up for the loss of information by yielding indirect estimates of the companion properties.

As an eccentric binary around an expanding red-giant star, it is interesting to explore the potential future evolution of the system. As the primary continues to ascend the red giant branch, the models used in Sec \ref{sec:param} predict that a ${\sim}1\Ms$ star reaches a radius of ${\sim}160\Rs$ (0.75AU) at the tip of the red giant branch. The current separation of the two stars at periastron is $R_{\textrm{peri}}=7.25\pm0.10\textrm{R}_{\star}$, or $0.19\pm0.01$AU, and as such, the two stars will meet during the ascent of the primary star up the giant branch. 

What configuration the orbit of the secondary will end up in is determined by the rate of tidal circularisation for the system. If the rate is high, then the orbital energy (and momentum) of the M-dwarf will be dissipated in the deep convective zone of the primary, as will the orbital eccentricity. To estimate the tidal circularisation timescale for a star with a convective envelope we used Eq 16 of \cite{1995Claret}, i.e.,
\begin{equation}
\tau_{\textrm{circ}}=(1.99\e{3}\textrm{yr})M^{3}\frac{(1+q)^{5/3}}{q}L^{-1/3}\lambda_{2}^{-1}\frac{P^{16/3}(\textrm{days})}{R^{22/3}}
\label{eq:tidecirc}
\end{equation}

In the above equation $M$, $R$ and $L$ are the total stellar mass, radius and luminosity in solar units, $q$ is the mass ratio of the components ($M_{\star}/M_{2}$). $\lambda_{2}$ is known as the tidal constant and is related to the internal structure of the star. We take a representative value from \cite{1995Claret} of $\lambda_{2}=0.006$ (see their Fig 3).

Assuming the orbital period, $P$, has remained constant throughout the history of the system, $\tau_{\textrm{circ}}$ during the main sequence lifetime would have been far greater than the expected main sequence lifetime of the red giant ($\tau_{\textrm{circ}}\approx 2\e{17}$yr). The main sequence luminosity and radius have been estimated using the relations $L\propto M^{3.9}$ and $R\propto M^{0.8}$, taking $M$ from Table \ref{tab:3890params}, neglecting any mass loss throughout the star's lifetime thus far.

As the primary ascends the red giant branch, $\tau_{\textrm{circ}}$ will dramatically decrease, due to the strong dependence on $R$ (and to a lesser extent, the dependence on $L$). At the present epoch $\tau_{\textrm{circ}}\sim9\times10^{11}$ Gyr. Only when the system is significantly more evolved does the radius dependence overcome the period dependence of Eq \ref{eq:tidecirc}. For instance, assuming $P$ does not change before $R_{\star}=R_{\textrm{peri}}$ then $\tau_{\textrm{circ}}\sim2\e{5}$yr when the stars come into contact. As the primary continues to evolve the M-dwarf may become embedded in the expanding envelope, leading to mass transfer between the stars in a common envelope phase. Additionally the strong drag forces on the secondary in such a configuration may lead to the ejection of the common envelope, and a significant decrease in the orbital period. Since the M-dwarf will encounter the expanding envelope before the primary has reached the tip of the RGB, the primary will not have gone through the helium flash. In the event that the common envelope is ejected, the helium core may be exposed as a sdB star, and the binary left in a close orbit, thus providing a mechanism for the formation of sdB stars. If the drag on the M dwarf is sufficient before the ejection of the envelope, the red giant core and the M-dwarf could collide or merge inside the common envelope.

\section*{Acknowledgements}
We would like to thank the referee for their helpful comments and suggestions. The authors thank Beno\^{i}t Mosser and Mathieu Vrard for providing data regarding the period spacing of the star, as well as Nathalie Theme\ss l for many helpful discussions. The research leading to the presented results has received funding from the European Research Council under the European Community's Seventh Framework Programme (FP7/2007-2013) / ERC grant agreement no. 338251 (StellarAges). This research has made use of the Exoplanet Follow-up Observation Program website, which is operated by the California Institute of Technology, under contract with the National Aeronautics and Space Administration under the Exoplanet Exploration Program. The authors acknowledge the support of the UK Science and Technology Facilities Council (STFC). Funding for the Stellar Astrophysics Centre is provided by the Danish National Research Foundation (Grant DNRF106). This research has made use of the NASA Exoplanet Archive, which is operated by the California Institute of Technology, under contract with the National Aeronautics and Space Administration under the Exoplanet Exploration Program. T. L. Campante acknowledges support from the European Union's Horizon 2020 research and innovation program under the Marie Sk\l{}odowska-Curie grant agreement No. 792848. A.M. acknowledges support from the ERC Consolidator Grant funding scheme (project ASTEROCHRONOMETRY, G.A. n. 772293).




\bibliographystyle{mnras}
\bibliography{biblio.bib} 

\begin{thebibliography}{}
\makeatletter
\relax
\def\mn@urlcharsother{\let\do\@makeother \do\$\do\&\do\#\do\^\do\_\do\%\do\~}
\def\mn@doi{\begingroup\mn@urlcharsother \@ifnextchar [ {\mn@doi@}
  {\mn@doi@[]}}
\def\mn@doi@[#1]#2{\def\@tempa{#1}\ifx\@tempa\@empty \href
  {http://dx.doi.org/#2} {doi:#2}\else \href {http://dx.doi.org/#2} {#1}\fi
  \endgroup}
\def\mn@eprint#1#2{\mn@eprint@#1:#2::\@nil}
\def\mn@eprint@arXiv#1{\href {http://arxiv.org/abs/#1} {{\tt arXiv:#1}}}
\def\mn@eprint@dblp#1{\href {http://dblp.uni-trier.de/rec/bibtex/#1.xml}
  {dblp:#1}}
\def\mn@eprint@#1:#2:#3:#4\@nil{\def\@tempa {#1}\def\@tempb {#2}\def\@tempc
  {#3}\ifx \@tempc \@empty \let \@tempc \@tempb \let \@tempb \@tempa \fi \ifx
  \@tempb \@empty \def\@tempb {arXiv}\fi \@ifundefined
  {mn@eprint@\@tempb}{\@tempb:\@tempc}{\expandafter \expandafter \csname
  mn@eprint@\@tempb\endcsname \expandafter{\@tempc}}}

\bibitem[\protect\citeauthoryear{{Aceituno} et~al.,}{{Aceituno}
  et~al.}{2013}]{2013Aceituno}
{Aceituno} J.,  et~al., 2013, \mn@doi [\aap] {10.1051/0004-6361/201220361},
  \href {http://adsabs.harvard.edu/abs/2013A%26A...552A..31A} {552, A31}

\bibitem[\protect\citeauthoryear{{Akeson} et~al.,}{{Akeson}
  et~al.}{2013}]{ExArc2013}
{Akeson} R.~L.,  et~al., 2013, \mn@doi [\pasp] {10.1086/672273}, \href
  {http://adsabs.harvard.edu/abs/2013PASP..125..989A} {125, 989}

\bibitem[\protect\citeauthoryear{{Albrecht}, {Reffert}, {Snellen},
  {Quirrenbach}  \& {Mitchell}}{{Albrecht} et~al.}{2007}]{2007A&A...474..565A}
{Albrecht} S.,  {Reffert} S.,  {Snellen} I.,  {Quirrenbach} A.,   {Mitchell}
  D.~S.,  2007, \mn@doi [\aap] {10.1051/0004-6361:20077953}, \href
  {http://adsabs.harvard.edu/abs/2007A%26A...474..565A} {474, 565}

\bibitem[\protect\citeauthoryear{{Albrecht}, {Reffert}, {Snellen}  \&
  {Winn}}{{Albrecht} et~al.}{2009}]{2009Natur.461..373A}
{Albrecht} S.,  {Reffert} S.,  {Snellen} I.~A.~G.,   {Winn} J.~N.,  2009,
  \mn@doi [\nat] {10.1038/nature08408}, \href
  {http://adsabs.harvard.edu/abs/2009Natur.461..373A} {461, 373}

\bibitem[\protect\citeauthoryear{{Albrecht}, {Winn}, {Carter}, {Snellen}  \&
  {de Mooij}}{{Albrecht} et~al.}{2011}]{2011ApJ...726...68A}
{Albrecht} S.,  {Winn} J.~N.,  {Carter} J.~A.,  {Snellen} I.~A.~G.,   {de
  Mooij} E.~J.~W.,  2011, \mn@doi [\apj] {10.1088/0004-637X/726/2/68}, \href
  {http://adsabs.harvard.edu/abs/2011ApJ...726...68A} {726, 68}

\bibitem[\protect\citeauthoryear{{Albrecht}, {Setiawan}, {Torres}, {Fabrycky}
  \& {Winn}}{{Albrecht} et~al.}{2013}]{2013ApJ...767...32A}
{Albrecht} S.,  {Setiawan} J.,  {Torres} G.,  {Fabrycky} D.~C.,   {Winn} J.~N.,
   2013, \mn@doi [\apj] {10.1088/0004-637X/767/1/32}, \href
  {http://adsabs.harvard.edu/abs/2013ApJ...767...32A} {767, 32}

\bibitem[\protect\citeauthoryear{{Albrecht} et~al.,}{{Albrecht}
  et~al.}{2014}]{2014ApJ...785...83A}
{Albrecht} S.,  et~al., 2014, \mn@doi [\apj] {10.1088/0004-637X/785/2/83},
  \href {http://adsabs.harvard.edu/abs/2014ApJ...785...83A} {785, 83}

\bibitem[\protect\citeauthoryear{{Arentoft} et~al.,}{{Arentoft}
  et~al.}{2008}]{2008ApJ...687.1180A}
{Arentoft} T.,  et~al., 2008, \mn@doi [\apj] {10.1086/592040}, \href
  {http://adsabs.harvard.edu/abs/2008ApJ...687.1180A} {687, 1180}

\bibitem[\protect\citeauthoryear{{Baglin} et~al.,}{{Baglin}
  et~al.}{2006}]{2006Baglin_corot}
{Baglin} A.,  et~al., 2006, in 36th COSPAR Scientific Assembly.

\bibitem[\protect\citeauthoryear{{Baraffe}, {Homeier}, {Allard}  \&
  {Chabrier}}{{Baraffe} et~al.}{2015}]{2015A&A...577A..42B}
{Baraffe} I.,  {Homeier} D.,  {Allard} F.,   {Chabrier} G.,  2015, \mn@doi
  [\aap] {10.1051/0004-6361/201425481}, \href
  {http://adsabs.harvard.edu/abs/2015A%26A...577A..42B} {577, A42}

\bibitem[\protect\citeauthoryear{{Bate}, {Lodato}  \& {Pringle}}{{Bate}
  et~al.}{2010}]{2010MNRAS.401.1505B}
{Bate} M.~R.,  {Lodato} G.,   {Pringle} J.~E.,  2010, \mn@doi [\mnras]
  {10.1111/j.1365-2966.2009.15773.x}, \href
  {http://adsabs.harvard.edu/abs/2010MNRAS.401.1505B} {401, 1505}

\bibitem[\protect\citeauthoryear{{Beck} et~al.,}{{Beck}
  et~al.}{2014a}]{2014A&A...564A..36B}
{Beck} P.~G.,  et~al., 2014a, \mn@doi [\aap] {10.1051/0004-6361/201322477},
  \href {http://adsabs.harvard.edu/abs/2014A%26A...564A..36B} {564, A36}

\bibitem[\protect\citeauthoryear{{Beck} et~al.,}{{Beck}
  et~al.}{2014b}]{2014Beck}
{Beck} P.~G.,  et~al., 2014b, \mn@doi [\aap] {10.1051/0004-6361/201322477},
  \href {http://adsabs.harvard.edu/abs/2014A%26A...564A..36B} {564, A36}

\bibitem[\protect\citeauthoryear{{Bedding} et~al.,}{{Bedding}
  et~al.}{2010}]{2010ApJ...713..935B}
{Bedding} T.~R.,  et~al., 2010, \mn@doi [\apj] {10.1088/0004-637X/713/2/935},
  \href {http://adsabs.harvard.edu/abs/2010ApJ...713..935B} {713, 935}

\bibitem[\protect\citeauthoryear{{Benbakoura}, {Gaulme}, {McKeever}, {Beck},
  {Jackiewicz}  \& {Garc{\'{\i}}a}}{{Benbakoura}
  et~al.}{2017}]{2017sf2a.conf...89B}
{Benbakoura} M.,  {Gaulme} P.,  {McKeever} J.,  {Beck} P.~G.,  {Jackiewicz} J.,
    {Garc{\'{\i}}a} R.~A.,  2017, in {Reyl{\'e}} C.,  {Di Matteo} P.,  {Herpin}
  F.,  {Lagadec} E.,  {Lan{\c c}on} A.,  {Meliani} Z.,   {Royer} F.,  eds,
  SF2A-2017: Proceedings of the Annual meeting of the French Society of
  Astronomy and Astrophysics. pp 89--92 (\mn@eprint {arXiv} {1712.01082})

\bibitem[\protect\citeauthoryear{{Birkby} et~al.,}{{Birkby}
  et~al.}{2012}]{2012MNRAS.426.1507B}
{Birkby} J.,  et~al., 2012, \mn@doi [\mnras]
  {10.1111/j.1365-2966.2012.21514.x}, \href
  {http://adsabs.harvard.edu/abs/2012MNRAS.426.1507B} {426, 1507}

\bibitem[\protect\citeauthoryear{{Bonnell}, {Arcoragi}, {Martel}  \&
  {Bastien}}{{Bonnell} et~al.}{1992}]{1992ApJ...400..579B}
{Bonnell} I.,  {Arcoragi} J.-P.,  {Martel} H.,   {Bastien} P.,  1992, \mn@doi
  [\apj] {10.1086/172020}, \href
  {http://adsabs.harvard.edu/abs/1992ApJ...400..579B} {400, 579}

\bibitem[\protect\citeauthoryear{{Borucki} et~al.,}{{Borucki}
  et~al.}{2010}]{2010Borucki_kepler}
{Borucki} W.~J.,  et~al., 2010, \mn@doi [Science] {10.1126/science.1185402},
  \href {http://adsabs.harvard.edu/abs/2010Sci...327..977B} {327, 977}

\bibitem[\protect\citeauthoryear{{Brogaard} et~al.,}{{Brogaard}
  et~al.}{2016}]{2016AN....337..793B}
{Brogaard} K.,  et~al., 2016, \mn@doi [Astronomische Nachrichten]
  {10.1002/asna.201612374}, \href
  {http://adsabs.harvard.edu/abs/2016AN....337..793B} {337, 793}

\bibitem[\protect\citeauthoryear{{Brogaard} et~al.,}{{Brogaard}
  et~al.}{2018}]{2018MNRAS.476.3729B}
{Brogaard} K.,  et~al., 2018, \mn@doi [\mnras] {10.1093/mnras/sty268}, \href
  {http://adsabs.harvard.edu/abs/2018MNRAS.476.3729B} {476, 3729}

\bibitem[\protect\citeauthoryear{{Brown}, {Gilliland}, {Noyes}  \&
  {Ramsey}}{{Brown} et~al.}{1991}]{1991ApJ...368..599B}
{Brown} T.~M.,  {Gilliland} R.~L.,  {Noyes} R.~W.,   {Ramsey} L.~W.,  1991,
  \mn@doi [\apj] {10.1086/169725}, \href
  {http://adsabs.harvard.edu/abs/1991ApJ...368..599B} {368, 599}

\bibitem[\protect\citeauthoryear{{Buchhave} et~al.,}{{Buchhave}
  et~al.}{2010}]{2010ApJ...720.1118B}
{Buchhave} L.~A.,  et~al., 2010, \mn@doi [\apj] {10.1088/0004-637X/720/2/1118},
  \href {http://adsabs.harvard.edu/abs/2010ApJ...720.1118B} {720, 1118}

\bibitem[\protect\citeauthoryear{{Buchhave} et~al.,}{{Buchhave}
  et~al.}{2012}]{2012Natur.486..375B}
{Buchhave} L.~A.,  et~al., 2012, \mn@doi [\nat] {10.1038/nature11121}, \href
  {http://adsabs.harvard.edu/abs/2012Natur.486..375B} {486, 375}

\bibitem[\protect\citeauthoryear{Burrows, Hubbard, Lunine  \& Liebert}{Burrows
  et~al.}{2001}]{RevModPhys.73.719}
Burrows A.,  Hubbard W.~B.,  Lunine J.~I.,   Liebert J.,  2001, \mn@doi [Rev.
  Mod. Phys.] {10.1103/RevModPhys.73.719}, 73, 719

\bibitem[\protect\citeauthoryear{{Campante} et~al.,}{{Campante}
  et~al.}{2016}]{2016ApJ...819...85C}
{Campante} T.~L.,  et~al., 2016, \mn@doi [\apj] {10.3847/0004-637X/819/1/85},
  \href {http://adsabs.harvard.edu/abs/2016ApJ...819...85C} {819, 85}

\bibitem[\protect\citeauthoryear{{Chabrier}, {Gallardo}  \&
  {Baraffe}}{{Chabrier} et~al.}{2007}]{2007A&A...472L..17C}
{Chabrier} G.,  {Gallardo} J.,   {Baraffe} I.,  2007, \mn@doi [\aap]
  {10.1051/0004-6361:20077702}, \href
  {http://adsabs.harvard.edu/abs/2007A%26A...472L..17C} {472, L17}

\bibitem[\protect\citeauthoryear{{Chaplin} et~al.,}{{Chaplin}
  et~al.}{2014}]{2014ApJS..210....1C}
{Chaplin} W.~J.,  et~al., 2014, \mn@doi [\apjs] {10.1088/0067-0049/210/1/1},
  \href {http://adsabs.harvard.edu/abs/2014ApJS..210....1C} {210, 1}

\bibitem[\protect\citeauthoryear{{Claret}, {Gimenez}  \& {Cunha}}{{Claret}
  et~al.}{1995}]{1995Claret}
{Claret} A.,  {Gimenez} A.,   {Cunha} N.~C.~S.,  1995, \aap, \href
  {http://adsabs.harvard.edu/abs/1995A%26A...299..724C} {299, 724}

\bibitem[\protect\citeauthoryear{{Davies} \& {Miglio}}{{Davies} \&
  {Miglio}}{2016}]{2016AN....337..774D}
{Davies} G.~R.,  {Miglio} A.,  2016, \mn@doi [Astronomische Nachrichten]
  {10.1002/asna.201612371}, \href
  {http://adsabs.harvard.edu/abs/2016AN....337..774D} {337, 774}

\bibitem[\protect\citeauthoryear{{De Ridder} et~al.,}{{De Ridder}
  et~al.}{2009}]{2009Natur.459..398D}
{De Ridder} J.,  et~al., 2009, \mn@doi [\nat] {10.1038/nature08022}, \href
  {http://adsabs.harvard.edu/abs/2009Natur.459..398D} {459, 398}

\bibitem[\protect\citeauthoryear{{Demory} et~al.,}{{Demory}
  et~al.}{2009}]{2009A&A...505..205D}
{Demory} B.-O.,  et~al., 2009, \mn@doi [\aap] {10.1051/0004-6361/200911976},
  \href {http://adsabs.harvard.edu/abs/2009A%26A...505..205D} {505, 205}

\bibitem[\protect\citeauthoryear{{Eggenberger} et~al.,}{{Eggenberger}
  et~al.}{2017}]{2017Eggenberger}
{Eggenberger} P.,  et~al., 2017, \mn@doi [\aap] {10.1051/0004-6361/201629459},
  \href {http://adsabs.harvard.edu/abs/2017A%26A...599A..18E} {599, A18}

\bibitem[\protect\citeauthoryear{{Fabrycky} \& {Winn}}{{Fabrycky} \&
  {Winn}}{2009}]{2009ApJ...696.1230F}
{Fabrycky} D.~C.,  {Winn} J.~N.,  2009, \mn@doi [\apj]
  {10.1088/0004-637X/696/2/1230}, \href
  {http://adsabs.harvard.edu/abs/2009ApJ...696.1230F} {696, 1230}

\bibitem[\protect\citeauthoryear{Feiden \& Chaboyer}{Feiden \&
  Chaboyer}{2012}]{0004-637X-757-1-42}
Feiden G.~A.,  Chaboyer B.,  2012, The Astrophysical Journal, 757, 42

\bibitem[\protect\citeauthoryear{{Foreman-Mackey}, {Hogg}, {Lang}  \&
  {Goodman}}{{Foreman-Mackey} et~al.}{2013}]{emcee}
{Foreman-Mackey} D.,  {Hogg} D.~W.,  {Lang} D.,   {Goodman} J.,  2013, \mn@doi
  [\pasp] {10.1086/670067}, \href
  {http://adsabs.harvard.edu/abs/2013PASP..125..306F} {125, 306}

\bibitem[\protect\citeauthoryear{{Frandsen} et~al.,}{{Frandsen}
  et~al.}{2013}]{2013A&A...556A.138F}
{Frandsen} S.,  et~al., 2013, \mn@doi [\aap] {10.1051/0004-6361/201321817},
  \href {http://adsabs.harvard.edu/abs/2013A%26A...556A.138F} {556, A138}

\bibitem[\protect\citeauthoryear{{Gaudi} \& {Winn}}{{Gaudi} \&
  {Winn}}{2007}]{2007ApJ...655..550G}
{Gaudi} B.~S.,  {Winn} J.~N.,  2007, \mn@doi [\apj] {10.1086/509910}, \href
  {http://adsabs.harvard.edu/abs/2007ApJ...655..550G} {655, 550}

\bibitem[\protect\citeauthoryear{{Gaulme}, {McKeever}, {Rawls}, {Jackiewicz},
  {Mosser}  \& {Guzik}}{{Gaulme} et~al.}{2013}]{2013Gaulme}
{Gaulme} P.,  {McKeever} J.,  {Rawls} M.~L.,  {Jackiewicz} J.,  {Mosser} B.,
  {Guzik} J.~A.,  2013, \mn@doi [\apj] {10.1088/0004-637X/767/1/82}, \href
  {http://adsabs.harvard.edu/abs/2013ApJ...767...82G} {767, 82}

\bibitem[\protect\citeauthoryear{{Gaulme}, {Jackiewicz}, {Appourchaux}  \&
  {Mosser}}{{Gaulme} et~al.}{2014}]{2014ApJ...785....5G}
{Gaulme} P.,  {Jackiewicz} J.,  {Appourchaux} T.,   {Mosser} B.,  2014, \mn@doi
  [\apj] {10.1088/0004-637X/785/1/5}, \href
  {http://adsabs.harvard.edu/abs/2014ApJ...785....5G} {785, 5}

\bibitem[\protect\citeauthoryear{{Gaulme} et~al.,}{{Gaulme}
  et~al.}{2016}]{2016Gaulme}
{Gaulme} P.,  et~al., 2016, \mn@doi [\apj] {10.3847/0004-637X/832/2/121}, \href
  {http://adsabs.harvard.edu/abs/2016ApJ...832..121G} {832, 121}

\bibitem[\protect\citeauthoryear{{Gehan}, {Mosser}, {Michel}, {Samadi}  \&
  {Kallinger}}{{Gehan} et~al.}{2018}]{2018A&A...616A..24G}
{Gehan} C.,  {Mosser} B.,  {Michel} E.,  {Samadi} R.,   {Kallinger} T.,  2018,
  \mn@doi [\aap] {10.1051/0004-6361/201832822}, \href
  {http://adsabs.harvard.edu/abs/2018A%26A...616A..24G} {616, A24}

\bibitem[\protect\citeauthoryear{Gizon \& Solanki}{Gizon \&
  Solanki}{2003}]{Gizon2003}
Gizon L.,  Solanki S.~K.,  2003, The Astrophysical Journal, 589, 1009

\bibitem[\protect\citeauthoryear{{Guo}, {Gies}  \& {Fuller}}{{Guo}
  et~al.}{2017}]{2017ApJ...834...59G}
{Guo} Z.,  {Gies} D.~R.,   {Fuller} J.,  2017, \mn@doi [\apj]
  {10.3847/1538-4357/834/1/59}, \href
  {http://adsabs.harvard.edu/abs/2017ApJ...834...59G} {834, 59}

\bibitem[\protect\citeauthoryear{{Hambleton} et~al.,}{{Hambleton}
  et~al.}{2018}]{2018MNRAS.473.5165H}
{Hambleton} K.,  et~al., 2018, \mn@doi [\mnras] {10.1093/mnras/stx2673}, \href
  {http://adsabs.harvard.edu/abs/2018MNRAS.473.5165H} {473, 5165}

\bibitem[\protect\citeauthoryear{{Hekker} et~al.,}{{Hekker}
  et~al.}{2009}]{2009A&A...506..465H}
{Hekker} S.,  et~al., 2009, \mn@doi [\aap] {10.1051/0004-6361/200911858}, \href
  {http://adsabs.harvard.edu/abs/2009A%26A...506..465H} {506, 465}

\bibitem[\protect\citeauthoryear{{Hekker} et~al.,}{{Hekker}
  et~al.}{2010}]{2010ApJ...713L.187H}
{Hekker} S.,  et~al., 2010, \mn@doi [\apjl] {10.1088/2041-8205/713/2/L187},
  \href {http://adsabs.harvard.edu/abs/2010ApJ...713L.187H} {713, L187}

\bibitem[\protect\citeauthoryear{{Hekker} et~al.,}{{Hekker}
  et~al.}{2011}]{2011MNRAS.414.2594H}
{Hekker} S.,  et~al., 2011, \mn@doi [\mnras]
  {10.1111/j.1365-2966.2011.18574.x}, \href
  {http://adsabs.harvard.edu/abs/2011MNRAS.414.2594H} {414, 2594}

\bibitem[\protect\citeauthoryear{{Hekker}, {Elsworth}  \& {Angelou}}{{Hekker}
  et~al.}{2018}]{2018A&A...610A..80H}
{Hekker} S.,  {Elsworth} Y.,   {Angelou} G.~C.,  2018, \mn@doi [\aap]
  {10.1051/0004-6361/201731264}, \href
  {http://adsabs.harvard.edu/abs/2018A%26A...610A..80H} {610, A80}

\bibitem[\protect\citeauthoryear{{Huber} et~al.,}{{Huber}
  et~al.}{2013}]{Kepler56}
{Huber} D.,  et~al., 2013, \mn@doi [Science] {10.1126/science.1242066}, \href
  {http://adsabs.harvard.edu/abs/2013Sci...342..331H} {342, 331}

\bibitem[\protect\citeauthoryear{{Kallinger} et~al.,}{{Kallinger}
  et~al.}{2014}]{Kallinger2014}
{Kallinger} T.,  et~al., 2014, \mn@doi [\aap] {10.1051/0004-6361/201424313},
  \href {http://adsabs.harvard.edu/abs/2014A%26A...570A..41K} {570, A41}

\bibitem[\protect\citeauthoryear{{Kaltenegger} \& {Traub}}{{Kaltenegger} \&
  {Traub}}{2009}]{Kaltenegger2009}
{Kaltenegger} L.,  {Traub} W.~A.,  2009, \mn@doi [\apj]
  {10.1088/0004-637X/698/1/519}, \href
  {http://adsabs.harvard.edu/abs/2009ApJ...698..519K} {698, 519}

\bibitem[\protect\citeauthoryear{{Kamiaka}, {Benomar}  \& {Suto}}{{Kamiaka}
  et~al.}{2018}]{2018arXiv180507044K}
{Kamiaka} S.,  {Benomar} O.,   {Suto} Y.,  2018, preprint, \href
  {http://adsabs.harvard.edu/abs/2018arXiv180507044K} {} (\mn@eprint {arXiv}
  {1805.07044})

\bibitem[\protect\citeauthoryear{{Kesseli}, {Muirhead}, {Mann}  \&
  {Mace}}{{Kesseli} et~al.}{2018}]{2018AJ....155..225K}
{Kesseli} A.~Y.,  {Muirhead} P.~S.,  {Mann} A.~W.,   {Mace} G.,  2018, \mn@doi
  [\aj] {10.3847/1538-3881/aabccb}, \href
  {http://adsabs.harvard.edu/abs/2018AJ....155..225K} {155, 225}

\bibitem[\protect\citeauthoryear{{Kiefer}, {Schad}, {Herzberg}  \&
  {Roth}}{{Kiefer} et~al.}{2015}]{2015A&A...578A..56K}
{Kiefer} R.,  {Schad} A.,  {Herzberg} W.,   {Roth} M.,  2015, \mn@doi [\aap]
  {10.1051/0004-6361/201425474}, \href
  {http://adsabs.harvard.edu/abs/2015A%26A...578A..56K} {578, A56}

\bibitem[\protect\citeauthoryear{{Kjeldsen} et~al.,}{{Kjeldsen}
  et~al.}{2005}]{2005ApJ...635.1281K}
{Kjeldsen} H.,  et~al., 2005, \mn@doi [\apj] {10.1086/497530}, \href
  {http://adsabs.harvard.edu/abs/2005ApJ...635.1281K} {635, 1281}

\bibitem[\protect\citeauthoryear{{Kov{\'a}cs}, {Zucker}  \&
  {Mazeh}}{{Kov{\'a}cs} et~al.}{2002}]{2002Kovacs}
{Kov{\'a}cs} G.,  {Zucker} S.,   {Mazeh} T.,  2002, \mn@doi [\aap]
  {10.1051/0004-6361:20020802}, \href
  {http://adsabs.harvard.edu/abs/2002A%26A...391..369K} {391, 369}

\bibitem[\protect\citeauthoryear{{Kreidberg}}{{Kreidberg}}{2015}]{Batman2015}
{Kreidberg} L.,  2015, \mn@doi [\pasp] {10.1086/683602}, \href
  {http://adsabs.harvard.edu/abs/2015PASP..127.1161K} {127, 1161}

\bibitem[\protect\citeauthoryear{{Kumar}, {Ao}  \& {Quataert}}{{Kumar}
  et~al.}{1995}]{1995Kumar}
{Kumar} P.,  {Ao} C.~O.,   {Quataert} E.~J.,  1995, \mn@doi [\apj]
  {10.1086/176055}, \href {http://adsabs.harvard.edu/abs/1995ApJ...449..294K}
  {449, 294}

\bibitem[\protect\citeauthoryear{{Kurucz}}{{Kurucz}}{1992}]{1992IAUS..149..225K}
{Kurucz} R.~L.,  1992, in {Barbuy} B.,  {Renzini} A.,  eds,  IAU Symposium Vol.
  149, The Stellar Populations of Galaxies. p.~225

\bibitem[\protect\citeauthoryear{{Lillo-Box}, {Barrado}, {Mancini}, {Henning},
  {Figueira}, {Ciceri}  \& {Santos}}{{Lillo-Box} et~al.}{2015}]{2015Lilobox}
{Lillo-Box} J.,  {Barrado} D.,  {Mancini} L.,  {Henning} T.,  {Figueira} P.,
  {Ciceri} S.,   {Santos} N.,  2015, \mn@doi [\aap]
  {10.1051/0004-6361/201425272}, \href
  {http://adsabs.harvard.edu/abs/2015A%26A...576A..88L} {576, A88}

\bibitem[\protect\citeauthoryear{{Lomb}}{{Lomb}}{1976}]{1976Ap&SS..39..447L}
{Lomb} N.~R.,  1976, \mn@doi [\apss] {10.1007/BF00648343}, \href
  {http://adsabs.harvard.edu/abs/1976Ap%26SS..39..447L} {39, 447}

\bibitem[\protect\citeauthoryear{{L{\'o}pez-Morales} \&
  {Ribas}}{{L{\'o}pez-Morales} \& {Ribas}}{2005}]{2005ApJ...631.1120L}
{L{\'o}pez-Morales} M.,  {Ribas} I.,  2005, \mn@doi [\apj] {10.1086/432680},
  \href {http://adsabs.harvard.edu/abs/2005ApJ...631.1120L} {631, 1120}

\bibitem[\protect\citeauthoryear{{Lund} et~al.,}{{Lund}
  et~al.}{2014}]{2014A&A...570A..54L}
{Lund} M.~N.,  et~al., 2014, \mn@doi [\aap] {10.1051/0004-6361/201424326},
  \href {http://adsabs.harvard.edu/abs/2014A%26A...570A..54L} {570, A54}

\bibitem[\protect\citeauthoryear{{Lund} et~al.,}{{Lund}
  et~al.}{2016}]{2016Lund_hyades}
{Lund} M.~N.,  et~al., 2016, \mn@doi [\mnras] {10.1093/mnras/stw2160}, \href
  {http://adsabs.harvard.edu/abs/2016MNRAS.463.2600L} {463, 2600}

\bibitem[\protect\citeauthoryear{MacDonald \& Mullan}{MacDonald \&
  Mullan}{2013}]{0004-637X-765-2-126}
MacDonald J.,  Mullan D.~J.,  2013, The Astrophysical Journal, 765, 126

\bibitem[\protect\citeauthoryear{{Mandel} \& {Agol}}{{Mandel} \&
  {Agol}}{2002}]{2002MandelAgol}
{Mandel} K.,  {Agol} E.,  2002, \mn@doi [\apjl] {10.1086/345520}, \href
  {http://adsabs.harvard.edu/abs/2002ApJ...580L.171M} {580, L171}

\bibitem[\protect\citeauthoryear{{Mann}, {Feiden}, {Gaidos}, {Boyajian}  \&
  {von Braun}}{{Mann} et~al.}{2015}]{2015ApJ...804...64M}
{Mann} A.~W.,  {Feiden} G.~A.,  {Gaidos} E.,  {Boyajian} T.,   {von Braun} K.,
  2015, \mn@doi [\apj] {10.1088/0004-637X/804/1/64}, \href
  {http://adsabs.harvard.edu/abs/2015ApJ...804...64M} {804, 64}

\bibitem[\protect\citeauthoryear{{Mathur}, {Garc{\'{\i}}a}, {Huber}, {Regulo},
  {Stello}, {Beck}, {Houmani}  \& {Salabert}}{{Mathur}
  et~al.}{2016}]{2016ApJ...827...50M}
{Mathur} S.,  {Garc{\'{\i}}a} R.~A.,  {Huber} D.,  {Regulo} C.,  {Stello} D.,
  {Beck} P.~G.,  {Houmani} K.,   {Salabert} D.,  2016, \mn@doi [\apj]
  {10.3847/0004-637X/827/1/50}, \href
  {http://adsabs.harvard.edu/abs/2016ApJ...827...50M} {827, 50}

\bibitem[\protect\citeauthoryear{{Mazeh} \& {Shaham}}{{Mazeh} \&
  {Shaham}}{1979}]{1979A&A....77..145M}
{Mazeh} T.,  {Shaham} J.,  1979, \aap, \href
  {http://adsabs.harvard.edu/abs/1979A%26A....77..145M} {77, 145}

\bibitem[\protect\citeauthoryear{{Miglio} et~al.,}{{Miglio}
  et~al.}{2013}]{2013MNRAS.429..423M}
{Miglio} A.,  et~al., 2013, \mn@doi [\mnras] {10.1093/mnras/sts345}, \href
  {http://adsabs.harvard.edu/abs/2013MNRAS.429..423M} {429, 423}

\bibitem[\protect\citeauthoryear{{Miglio}, {Chaplin}, {Farmer}, {Kolb},
  {Girardi}, {Elsworth}, {Appourchaux}  \& {Handberg}}{{Miglio}
  et~al.}{2014}]{2014Miglio}
{Miglio} A.,  {Chaplin} W.~J.,  {Farmer} R.,  {Kolb} U.,  {Girardi} L.,
  {Elsworth} Y.,  {Appourchaux} T.,   {Handberg} R.,  2014, \mn@doi [\apjl]
  {10.1088/2041-8205/784/1/L3}, \href
  {http://adsabs.harvard.edu/abs/2014ApJ...784L...3M} {784, L3}

\bibitem[\protect\citeauthoryear{{Montalb{\'a}n}, {Miglio}, {Noels}, {Dupret},
  {Scuflaire}  \& {Ventura}}{{Montalb{\'a}n} et~al.}{2013}]{2013Mont}
{Montalb{\'a}n} J.,  {Miglio} A.,  {Noels} A.,  {Dupret} M.-A.,  {Scuflaire}
  R.,   {Ventura} P.,  2013, \mn@doi [\apj] {10.1088/0004-637X/766/2/118},
  \href {http://adsabs.harvard.edu/abs/2013ApJ...766..118M} {766, 118}

\bibitem[\protect\citeauthoryear{{Morton} \& {Winn}}{{Morton} \&
  {Winn}}{2014}]{2014MortonWinn}
{Morton} T.~D.,  {Winn} J.~N.,  2014, \mn@doi [\apj]
  {10.1088/0004-637X/796/1/47}, \href
  {http://adsabs.harvard.edu/abs/2014ApJ...796...47M} {796, 47}

\bibitem[\protect\citeauthoryear{{Mosser} et~al.,}{{Mosser}
  et~al.}{2012}]{2012Mosser_spindown}
{Mosser} B.,  et~al., 2012, \mn@doi [\aap] {10.1051/0004-6361/201220106}, \href
  {http://adsabs.harvard.edu/abs/2012A%26A...548A..10M} {548, A10}

\bibitem[\protect\citeauthoryear{{Mosser}, {Gehan}, {Belkacem}, {Samadi},
  {Michel}  \& {Goupil}}{{Mosser} et~al.}{2018}]{2018arXiv180708301M}
{Mosser} B.,  {Gehan} C.,  {Belkacem} K.,  {Samadi} R.,  {Michel} E.,
  {Goupil} M.,  2018, preprint, \href
  {http://adsabs.harvard.edu/abs/2018arXiv180708301M} {} (\mn@eprint {arXiv}
  {1807.08301})

\bibitem[\protect\citeauthoryear{{Parsons} et~al.,}{{Parsons}
  et~al.}{2018}]{2018MNRAS.tmp.2233P}
{Parsons} S.~G.,  et~al., 2018, \mn@doi [\mnras] {10.1093/mnras/sty2345}, \href
  {http://adsabs.harvard.edu/abs/2018MNRAS.tmp.2233P} {}

\bibitem[\protect\citeauthoryear{{Penoyre} \& {Stone}}{{Penoyre} \&
  {Stone}}{2018}]{2018arXiv180305917P}
{Penoyre} Z.,  {Stone} N.~C.,  2018, preprint, \href
  {http://adsabs.harvard.edu/abs/2018arXiv180305917P} {} (\mn@eprint {arXiv}
  {1803.05917})

\bibitem[\protect\citeauthoryear{{Pesnell}}{{Pesnell}}{1985}]{1985ApJ...292..238P}
{Pesnell} W.~D.,  1985, \mn@doi [\apj] {10.1086/163153}, \href
  {http://adsabs.harvard.edu/abs/1985ApJ...292..238P} {292, 238}

\bibitem[\protect\citeauthoryear{{Rawls} et~al.,}{{Rawls}
  et~al.}{2016}]{2016ApJ...818..108R}
{Rawls} M.~L.,  et~al., 2016, \mn@doi [\apj] {10.3847/0004-637X/818/2/108},
  \href {http://adsabs.harvard.edu/abs/2016ApJ...818..108R} {818, 108}

\bibitem[\protect\citeauthoryear{{Rodrigues} et~al.,}{{Rodrigues}
  et~al.}{2014}]{2014Rodrigues}
{Rodrigues} T.~S.,  et~al., 2014, \mn@doi [\mnras] {10.1093/mnras/stu1907},
  \href {http://adsabs.harvard.edu/abs/2014MNRAS.445.2758R} {445, 2758}

\bibitem[\protect\citeauthoryear{{Rodrigues} et~al.,}{{Rodrigues}
  et~al.}{2017}]{2017Rod}
{Rodrigues} T.~S.,  et~al., 2017, \mn@doi [\mnras] {10.1093/mnras/stx120},
  \href {http://adsabs.harvard.edu/abs/2017MNRAS.467.1433R} {467, 1433}

\bibitem[\protect\citeauthoryear{{Scargle}}{{Scargle}}{1982}]{1982ApJ...263..835S}
{Scargle} J.~D.,  1982, \mn@doi [\apj] {10.1086/160554}, \href
  {http://adsabs.harvard.edu/abs/1982ApJ...263..835S} {263, 835}

\bibitem[\protect\citeauthoryear{{Shporer} et~al.,}{{Shporer}
  et~al.}{2016}]{2016Shporer}
{Shporer} A.,  et~al., 2016, \mn@doi [\apj] {10.3847/0004-637X/829/1/34}, \href
  {http://adsabs.harvard.edu/abs/2016ApJ...829...34S} {829, 34}

\bibitem[\protect\citeauthoryear{{Silva Aguirre} et~al.,}{{Silva Aguirre}
  et~al.}{2018}]{2018MNRAS.475.5487S}
{Silva Aguirre} V.,  et~al., 2018, \mn@doi [\mnras] {10.1093/mnras/sty150},
  \href {http://adsabs.harvard.edu/abs/2018MNRAS.475.5487S} {475, 5487}

\bibitem[\protect\citeauthoryear{{Sing}}{{Sing}}{2010}]{2010Sing}
{Sing} D.~K.,  2010, \mn@doi [\aap] {10.1051/0004-6361/200913675}, \href
  {http://adsabs.harvard.edu/abs/2010A%26A...510A..21S} {510, A21}

\bibitem[\protect\citeauthoryear{{Smith}, {McMillan}  \& {Merline}}{{Smith}
  et~al.}{1987}]{1987ApJ...317L..79S}
{Smith} P.~H.,  {McMillan} R.~S.,   {Merline} W.~J.,  1987, \mn@doi [\apjl]
  {10.1086/184916}, \href {http://adsabs.harvard.edu/abs/1987ApJ...317L..79S}
  {317, L79}

\bibitem[\protect\citeauthoryear{{Soderblom}}{{Soderblom}}{2010}]{2010ARA&A..48..581S}
{Soderblom} D.~R.,  2010, \mn@doi [\araa]
  {10.1146/annurev-astro-081309-130806}, \href
  {http://adsabs.harvard.edu/abs/2010ARA%26A..48..581S} {48, 581}

\bibitem[\protect\citeauthoryear{{Szentgyorgyi} \& {Fur{\'e}sz}}{{Szentgyorgyi}
  \& {Fur{\'e}sz}}{2007}]{TRES}
{Szentgyorgyi} A.~H.,  {Fur{\'e}sz} G.,  2007, in {Kurtz} S.,  ed.,  Revista
  Mexicana de Astronomia y Astrofisica Conference Series Vol. 28, Revista
  Mexicana de Astronomia y Astrofisica Conference Series. pp 129--133

\bibitem[\protect\citeauthoryear{{Tassoul}}{{Tassoul}}{1980}]{1980Tassoul}
{Tassoul} M.,  1980, \mn@doi [\apjs] {10.1086/190678}, \href
  {http://adsabs.harvard.edu/abs/1980ApJS...43..469T} {43, 469}

\bibitem[\protect\citeauthoryear{{Theme{\ss}l} et~al.,}{{Theme{\ss}l}
  et~al.}{2018}]{2018MNRAS.tmp.1060T}
{Theme{\ss}l} N.,  et~al., 2018, \mn@doi [\mnras] {10.1093/mnras/sty1113},
  \href {http://adsabs.harvard.edu/abs/2018MNRAS.tmp.1060T} {}

\bibitem[\protect\citeauthoryear{{Thompson} et~al.,}{{Thompson}
  et~al.}{2012a}]{2012ApJ...753...86T}
{Thompson} S.~E.,  et~al., 2012a, \mn@doi [\apj] {10.1088/0004-637X/753/1/86},
  \href {http://adsabs.harvard.edu/abs/2012ApJ...753...86T} {753, 86}

\bibitem[\protect\citeauthoryear{{Thompson} et~al.,}{{Thompson}
  et~al.}{2012b}]{2012Thompson}
{Thompson} S.~E.,  et~al., 2012b, \mn@doi [\apj] {10.1088/0004-637X/753/1/86},
  \href {http://adsabs.harvard.edu/abs/2012ApJ...753...86T} {753, 86}

\bibitem[\protect\citeauthoryear{{Ulrich}}{{Ulrich}}{1986}]{1986ApJ...306L..37U}
{Ulrich} R.~K.,  1986, \mn@doi [\apjl] {10.1086/184700}, \href
  {http://adsabs.harvard.edu/abs/1986ApJ...306L..37U} {306, L37}

\bibitem[\protect\citeauthoryear{{VanderPlas} \& {Ivezi{\'c}}}{{VanderPlas} \&
  {Ivezi{\'c}}}{2015}]{2015ApJ...812...18V}
{VanderPlas} J.~T.,  {Ivezi{\'c}} {\v Z}.,  2015, \mn@doi [\apj]
  {10.1088/0004-637X/812/1/18}, \href
  {http://adsabs.harvard.edu/abs/2015ApJ...812...18V} {812, 18}

\bibitem[\protect\citeauthoryear{Vanderplas}{Vanderplas}{2015}]{jake_vanderplas_2015_14833}
Vanderplas J.,  2015, {gatspy: General tools for Astronomical Time Series in
  Python}, \mn@doi{10.5281/zenodo.14833}, \url
  {https://doi.org/10.5281/zenodo.14833}

\bibitem[\protect\citeauthoryear{{Vrard}, {Mosser}  \& {Samadi}}{{Vrard}
  et~al.}{2016}]{Vrard2016}
{Vrard} M.,  {Mosser} B.,   {Samadi} R.,  2016, \mn@doi [\aap]
  {10.1051/0004-6361/201527259}, \href
  {http://cdsads.u-strasbg.fr/abs/2016A%26A...588A..87V} {588, A87}

\bibitem[\protect\citeauthoryear{{Welsh} et~al.,}{{Welsh}
  et~al.}{2011}]{2011Welsh}
{Welsh} W.~F.,  et~al., 2011, \mn@doi [\apjs] {10.1088/0067-0049/197/1/4},
  \href {http://adsabs.harvard.edu/abs/2011ApJS..197....4W} {197, 4}

\bibitem[\protect\citeauthoryear{{Winn}}{{Winn}}{2007}]{2007ASPC..366..170W}
{Winn} J.~N.,  2007, in {Afonso} C.,  {Weldrake} D.,   {Henning} T.,  eds,
  Astronomical Society of the Pacific Conference Series Vol. 366, Transiting
  Extrapolar Planets Workshop. p.~170 (\mn@eprint {} {astro-ph/0612744})

\bibitem[\protect\citeauthoryear{{Winn}}{{Winn}}{2010a}]{Winn2010}
{Winn} J.~N.,  2010a, preprint (\mn@eprint {arXiv} {1001.2010})

\bibitem[\protect\citeauthoryear{{Winn}}{{Winn}}{2010b}]{2010arXiv1001.2010W}
{Winn} J.~N.,  2010b, preprint, \href
  {http://adsabs.harvard.edu/abs/2010arXiv1001.2010W} {} (\mn@eprint {arXiv}
  {1001.2010})

\bibitem[\protect\citeauthoryear{{Yu}, {Huber}, {Bedding}, {Stello}, {Hon},
  {Murphy}  \& {Khanna}}{{Yu} et~al.}{2018}]{2018ApJS..236...42Y}
{Yu} J.,  {Huber} D.,  {Bedding} T.~R.,  {Stello} D.,  {Hon} M.,  {Murphy}
  S.~J.,   {Khanna} S.,  2018, \mn@doi [\apjs] {10.3847/1538-4365/aaaf74},
  \href {http://adsabs.harvard.edu/abs/2018ApJS..236...42Y} {236, 42}

\bibitem[\protect\citeauthoryear{{da Silva} et~al.,}{{da Silva}
  et~al.}{2006}]{2006dasilva}
{da Silva} L.,  et~al., 2006, \mn@doi [\aap] {10.1051/0004-6361:20065105},
  \href {http://adsabs.harvard.edu/abs/2006A%26A...458..609D} {458, 609}

\makeatother
\end{thebibliography}

\bsp	
\label{lastpage}
\end{document}